\documentclass[12pt]{article}
\usepackage{amsmath,amssymb,amsfonts}
\usepackage{amsthm} 
\usepackage{graphicx}
\usepackage{enumerate}
\usepackage{natbib}
\usepackage{url} 

\usepackage{times}
\usepackage{bm}
\usepackage{xcolor}
\usepackage{comment}
\newcommand{\blind}{1}

\newtheorem{corollary}{Corollary}
\newtheorem{theorem}{Theorem}
\newtheorem{lemma}{Lemma}
\newtheorem{definition}{Definition}

\newtheorem{condition}{Condition}
\newtheorem{remark}{Remark}

\def\proj{\textup{proj}}
\def\res{\textup{res}}
\def\T{{ \mathrm{\scriptscriptstyle T} }}

\def\rs{\rightsquigarrow}
\def\ms{\mathcal S}
\def\sumw{\sum_{w=1}^{W}}
\def\sums{\sum_{s=1}^{M_w}}
\def\sumws{\sum_{ws \in \mathcal{S}}}

\def\alphaw{\alpha_w}
\def\HT{\textnormal{ht}}
\def\haj{\textnormal{haj}}
\def\tauht{\hat \tau_{\HT}}
\def\tauhaj{\hat \tau_{\haj}}
\def\tauxht{\hat \tau_{\HT,x}}
\def\tauxhaj{\hat \tau_{\haj,x}}
\def\cov{\textnormal{cov}}

\def\pr{\textnormal{pr}}

\def\cova{\textnormal{cov}_\textup{a}}

\def\pra{\textnormal{pr}_\textup{a}}

\def\ag{\textnormal{ag}}
\def\wls{\textnormal{wls}}
\def\L{\textsc{l}}
\def\tauhtL{\hat \tau_{\HT, \L}}
\def\tauhajL{\hat \tau_{\haj, \L}}

\def\tba{\hat \beta_\ag}
\def\tva{\hat V_\ag}
\def\tbw{\hat \beta_\wls}
\def\tvw{\hat V_\wls}

\def\tbal{\hat \beta_{\ag, \L}}
\def\tval{\hat V_{\ag, \L}}
\def\tbwl{\hat \beta_{\wls, \L}}
\def\tvwl{\hat V_{\wls, \L}}

\def\P{\textsc{p}}

\def\w{\textnormal{in}}
\def\b{\textnormal{b}}

\def\red{\color{red}}
\def\blue{\color{blue}}

\def\td{ {\tiny\dotfill}}
\def\itemc{\item[$\cdot$]}
\def\begina{\begin{eqnarray*}}
\def\enda{\end{eqnarray*}}
\def\beginy{\begin{eqnarray}}
\def\endy{\end{eqnarray}}
\def\mt{\mathcal {T}}
\def\htau{\hat\tau}
\def\hts{\htau_*}
\def\begini{\begin{itemize}}
\def\endi{\end{itemize}}
\def\lin{\textsc{l}}

\begin{document}

\def\spacingset#1{\renewcommand{\baselinestretch}%
{#1}\small\normalsize} \spacingset{1}


\if1\blind
{
  \title{\bf Rerandomization and covariate adjustment in split-plot designs}
  \author{Wenqi Shi\\
    Department of Industrial Engineering, Tsinghua University, \\
    Beijing, 100084, China\vspace{.2cm}\\
    Anqi Zhao\thanks{Zhao was supported by the Start-Up grant R-155-000-216-133 from the National University of Singapore.}\\
    Department of Statistics and Data Science, National University of Singapore, \\
    117546, Singapore\vspace{.2cm}\\
    Hanzhong Liu\thanks{Liu was supported by the National Natural Science Foundation of China (12071242).}\hspace{.2cm}\\
    Center for Statistical Science, Department of Industrial Engineering, \\
    Tsinghua University, Beijing, 100084, China}
    \date{}
  \maketitle
} \fi

\if0\blind
{
  \bigskip
  \bigskip
  \bigskip
  \begin{center}
    {\LARGE\bf Rerandomization and covariate adjustment in split-plot designs}
\end{center}
  \medskip
} \fi

\bigskip
\begin{abstract}
The split-plot design arises from agricultural sciences with experimental units, also known as subplots, nested within groups known as whole plots. 
It assigns the whole-plot intervention by a cluster randomization at the whole-plot level and assigns the subplot intervention by a stratified randomization at the subplot level. 
The randomization mechanism guarantees covariate balance on average at both the whole-plot and subplot levels, and ensures consistent inference of the average treatment effects by the Horvitz--Thompson and Hajek estimators. 
However, covariate imbalance often occurs in finite samples and subjects subsequent inference to possibly large variability and conditional bias. Rerandomization is widely used in the design stage of randomized experiments to improve covariate balance. The existing literature on rerandomization nevertheless focuses on designs with treatments assigned at either the unit or the group level, but not both, leaving the corresponding theory for rerandomization in split-plot designs an open problem. To fill the gap, we propose two strategies for conducting rerandomization in split-plot designs based on the Mahalanobis distance and establish the corresponding design-based theory. We show that rerandomization can improve the asymptotic efficiency of the Horvitz--Thompson and Hajek estimators.
Moreover, we propose two covariate adjustment methods in the analysis stage, 
which can further improve the asymptotic efficiency when combined with rerandomization.
The validity and improved efficiency of the proposed methods are demonstrated through numerical studies.
\end{abstract}

\noindent%
\vfill

\newpage
\spacingset{1.9} 

\section{Introduction}
The split-plot design has been widely used in agricultural sciences \citep{fisher1925statistical,Yates1937} and industrial experiments \citep{yates1935complex, jones}, and is gaining increasing popularity in social and biomedical sciences \citep{olken2007monitoring, meon, breza}. 
The experimental units, also known as the {\it subplots}, are nested within groups known as the {\it whole plots}. 
The split-plot design assigns the whole-plot intervention at the whole-plot level via a cluster randomization, and the subplot intervention at the subplot level via a stratified randomization. 
By design, subplots within the same whole plot receive the same level of the whole-plot intervention. This provides a convenient way to accommodate hard-to-change factors and avoid interference within whole plots.

\cite{kempthorne1952design} initiated the discussion on design-based inference 
of split-plot designs under the assumption of additive treatment effects. 
\cite{zhao2018randomization} loosened the requirement on additivity, and established the theory for finite-sample exact inference in {\it uniform} split-plot designs, i.e., the whole-plot sizes and proportions of treated units for the subplot intervention within each whole plot are constant across whole plots.
\cite{mukerjee2019causal} extended the discussion to possibly nonuniform split-plot designs, and established  the finite-sample exact theory for the Horvitz--Thompson estimator.
\cite{zhao2021reconciling} extended the theory to the Hajek estimator and established the consistency and asymptotic normality of the Horvitz--Thompson and Hajek estimators in possibly nonuniform split-plot designs.

In split-plot designs, experimenters often collect baseline covariates  at both the whole-plot and subplot levels. 
For example, in a split-plot design with students as subplots nested within whole plots of classes, class characteristics such as class size and teacher experience are whole-plot covariates, whereas student characteristics like race and gender are subplot covariates.
These baseline covariates are measured prior to the physical implementation of treatment assignments and hence not affected by the treatment. Randomization ensures that covariates are balanced across treatment levels on average. 
However, covariate imbalance often exists in a particular treatment allocation, and can complicate the interpretation of the experimental results   \citep{rubin2008comment,Morgan2012,krieger2019bka}. 
Rerandomization arose in such context and enforces covariate balance 
in the design stage of randomized experiments \citep{Morgan2012}. 
It has drawn much attention in the field of experimental design recently 
and is shown to ensure efficiency gains in various settings \citep[see, e.g.,][]{moulton2004covariate,Morgan2015,Li2018,Li2020factorial,Wang2021,Zhu2021,zdfactx,zdrep,Lu2022}.

The existing literature of rerandomization focuses on treatments assigned at either the unit or the group level, but not both, leaving the corresponding theory for rerandomization in split-plot designs an open problem. 
To fill this gap, we define {\it split-plot rerandomization} as a split-plot design compounded with rerandomization to balance covariates, and propose two split-plot rerandomization schemes based on the Mahalanobis distances of the Horvitz--Thompson and Hajek estimators of contrasts of covariate means to the origin, respectively. 
We derive the asymptotic distributions of the Horvitz--Thompson and Hajek estimators for the average treatment effects under split-plot rerandomization and demonstrate the efficiency gains relative to split-plot randomization. 

Regression adjustment 
 is another approach to dealing with covariate imbalance, taking place in the analysis stage. 
The existing literature sees efficiency gains by regression adjustment in various randomized experiments, including completely randomized experiments  \citep{lin2013,bloniarz2015lasso,Lei2020,zdfactx},
stratified randomized experiments \citep{Liu2019,Zhu2021block,Liu2022,Ma2022}, cluster randomized experiments \citep{Su2021,Lu2022}, 
completely or stratified randomized factorial experiments \citep{lu2016covariate,lu2016randomization,Liu2021,zdfact}, and split-plot experiments \citep{zhao2021reconciling}. 
In particular, \cite{zhao2021reconciling} studied several specifications for regression adjustment in split-plot designs 
and recommended an aggregate specification with full treatment-covariate interactions to ensure
efficiency gains when only whole-plot covariates are used.
Recent work by \cite{Li2020}, \cite{Wang2021}, and \cite{zdfactx, zdrep} further recommended combining rerandomization and regression adjustment in randomized experiments with treatments assigned at the unit level. 

In this paper, we propose a novel alternative to regression adjustment for covariate adjustment in the analysis stage, and provide a design-based theory for the combination of rerandomization and covariate adjustment in split-plot designs. 
We consider two  strategies for covariate adjustment in the analysis stage 
and derive their asymptotic distributions under split-plot rerandomization. 
The first strategy 
follows the regression formulation by \cite{zhao2021reconciling}, and ensures efficiency gains when only whole-plot covariates are used. 
Different from rerandomization with treatments assigned at the unit level, the asymptotic distributions of the regression-adjusted estimators under split-plot rerandomization are not normal, but convolutions of a normal component and a truncated normal component. 
Moreover, the regression adjustment 
may degrade efficiency if heterogeneous sub-plot covariates are used. 
The second strategy is new, and approaches covariate adjustment from a projection or conditional inference perspective. 
It adjusts an estimator for its asymptotic conditional bias given contrasts of covariate means, and yields estimators that are consistent and asymptotically normal under split-plot rerandomization with guaranteed efficiency gains.

We use the following notation. 
Let $\mathcal{I}(\cdot)$ be the indicator function. Let $\chi_n^2$ denote the chi-squared distribution with $n$ degrees of freedom.  
Let $0_m$ and $0_{m \times n} $ be the $m \times 1$ vector and $m \times n$ matrix of zeros, respectively. Let $1_m$ and $1_{m \times n} $ be the $m \times 1$ vector and $m \times n$ matrix of ones, respectively. Let $I_m$ be the $m \times m$ identity matrix. We suppress the dimensions when no confusion arises.
Let $\otimes$ and $\circ$ denote the Kronecker and Hadamard products of matrices, respectively. For two matrices $D_1$ and $D_2$, write $D_1 \geq D_2$ if $D_1 - D_2 $ is positive semi-definite. 
Let  $||\cdot||_\infty$ denote the $\ell_\infty$ norm. 
Let $\rs$ denote convergence in distribution. For a sequence of random variables $(U_n)_{n=1}^\infty$, write  $U_n \rs  U$ if as $n$ goes to $\infty$, the asymptotic distribution of $U_n$ equals the distribution of $U$.
Let $\pra$ and $\cova$ denote the asymptotic probability and covariance, respectively.

\section{Review of the split-plot design}

We follow the framework and notation in \cite{zhao2021reconciling}. 
Consider a $2^2$ split-plot design with two binary factors of interest, indexed by A, B $\in \{0,1\}$. 
This defines four treatment combinations, $\mathcal{T} = \{z = (a,b) :a,b=0,1\}$, where $a$ and $b$ index the levels of factors A and B, respectively. 
We abbreviate $(a,b)$ as $(ab)$ when no confusion would arise. 
Assume a study population of $N$ units nested in $W$ groups of possibly different sizes $M_w$ ($w = 1, \ldots,W$; $\sumw M_w = N$). 
We refer to each group as a whole plot and each unit as a subplot. 
Index by $ws$ the $s$th subplot in whole plot $w$, and let $\mathcal{S} = \{ws: w=1,\ldots,W;\ s=1,\ldots,M_w\}$ denote the entire population. 
The $2^2$ split-plot design assigns the units to different treatment combinations in two stages:
\begin{itemize}
    \item[(I)] the first stage assigns factor A  at the whole-plot level by a cluster randomization; that is,  it randomly assigns $W_a$ whole plots to receive level $a\in\{0,1\}$ of factor A for prespecified $W_a$'s with $W_0 + W_1 = W$;
    \item[(II)] the second stage assigns factor B at the subplot level by a stratified randomization; that is, it randomly assigns $M_{wb}$ units in whole plot $w$ to receive level $b\in\{0,1\}$ of factor B  for prespecified $M_{wb}$'s with $M_{w0} + M_{w1} = M_w, w=1,\ldots,M$, and the assignments across different whole plots are independent.
\end{itemize}
The final treatment of subplot $ws$, denoted by $Z_{ws} \in \mathcal T$, is then a combination of the level of factor A received by whole plot $w$ in stage (I) and the level of factor B received by itself in stage (II).  
Refer to factor A and factor B as the whole-plot and subplot factors, respectively. 
The probability of a whole plot assigned to level $a$ of factor A is $p_a=W_a/W$ for $a=0, 1$.
The probability of a subplot in whole plot $w$ assigned to level $b$ of factor B is $q_{wb}=M_{wb}/M_w$ for $w=1,\ldots,W$ and $b=0, 1$. 
Assume that the cluster and stratified randomizations are independent throughout. The probability of subplot $ws$ assigned to treatment $z=(ab)$ is $p_{ws}(z)=p_a q_{wb}$. 

Let $\bar M = N/W$ denote the average size of the whole plots, and let $\alphaw = M_w/\bar M$ denote the whole-plot size factor with  $W^{-1} \sumw\alphaw = 1$. We call a split-plot design {\it uniform} if $M_w$ and $M_{wb}$ are constants across $w = 1, \ldots, W$. A uniform design has $\alphaw = 1$ for all $w$.

We define treatment effects using the potential outcomes framework \citep{Neyman:1923,Rubin:1974}. Denote by $Y_{ws}(z)$ the potential outcome of subplot $ws$ if assigned to treatment $z \in \mt$, and let $\bar Y(z) = N^{-1} \sum_{ws \in \mathcal{S}} Y_{ws} (z)$ be the finite population average. 
%
%
%
The main effects and interaction under the $2^2$ split-plot design are  
\begin{align}
    \tau_{\textsc{a}} & = 2^{-1}\left\{\bar Y(10) + \bar Y(11) \right\}
    - 2^{-1}\left\{\bar Y(00) + \bar Y(01) \right\}, \notag \\
    \tau_{\textsc{b}} & = 2^{-1}\left\{\bar Y(01) + \bar Y(11) \right\}
    - 2^{-1}\left\{\bar Y(00) + \bar Y(10) \right\}, \notag \\
    \tau_{\textsc{ab}} & = \left\{\bar Y(00) + \bar Y(11) \right\}
    - \left\{\bar Y(01) + \bar Y(10) \right\} \notag 
\label{Def tau}
\end{align}
\citep{mukerjee2019causal,zhao2021reconciling}. 
Let $\bar{Y} = (\bar{Y}(00), \bar{Y}(01), \bar{Y}(10),\bar{Y}(11))^\T$ vectorize the $\bar Y(z)$'s in lexicographical order of $z$.  
We write the three effects in vector form as 
\begina
\tau =(\tau_{\textsc{a}}, \tau_{\textsc{b}}, \tau_{\textsc{ab}})^\T= G \bar Y
\enda with  $G = (g_{\textsc{a}}, g_{\textsc{b}}, g_{\textsc{ab}})^\T$ and $g_{\textsc{a}} = 2^{-1}(-1, -1, 1, 1)^\T$, $g_{\textsc{b}} = 2^{-1}(-1, 1, -1, 1)^\T$, $g_{\textsc{ab}} = (1, -1, -1, 1)^\T$.
There are other effects of interest, $\tau_g = g^\T \bar Y$, where $g$ is a $4\times 1$ contrast vector with $g^\T 1_4 = 0$ \citep{Cuesta2019,zdfact}. Such a $g$ can be represented by a linear combination of $g_{\textsc{a}}$, $g_{\textsc{b}}$, and $g_{\textsc{ab}}$ such that $\tau_g$ is a linear transformation of $\tau$. To simplify the presentation, we focus on $\tau$ in this paper. 

The observed outcome for subplot $ws$ is $Y_{ws} = \sum_{z \in \mathcal{T} } \mathcal{I}(Z_{ws}=z) Y_{ws}(z)$. 
Let $\mathcal{S}(z) = \{ws: Z_{ws} = z, \ ws \in \mathcal{S} \}$ denote the set of subplots assigned to treatment $z \in \mathcal{T}$.
The Horvitz--Thompson estimator for $\bar{Y}(z)$ is
\begina
\hat Y_{\HT} (z) = N^{-1}\sum_{ws \in \mathcal{S}(z)}{p^{-1}_{ws}}(z){Y_{ws}} = N^{-1}\sum_{ws \in \mathcal{S}}\frac{\mathcal{I}(Z_{ws}=z)}{p_{ws}(z)}Y_{ws}(z),
\enda
{and is unbiased under the $2^2$ split-plot randomization.}
Let $\hat Y_{\HT}$ be the vectorization of $\{ \hat Y_{\HT} (z) \}_{z \in \mathcal{T}}$ in lexicographical order of $z$. 
We call $\tauht = G \hat Y_{\HT}$ the Horvitz--Thompson estimator of $\tau$, which is unbiased under the split-plot randomization.
A major drawback of the Horvitz--Thompson estimator is that it is not invariant to location shifts \citep{Fuller2009}. To address this issue, another widely used estimator, the Hajek estimator, is defined as 
$$
 \hat Y_{\haj} (z)=\frac{\hat Y_{\HT}(z)}{\hat 1_{\HT}(z)},
$$
where $\hat 1_{\HT}(z) = N^{-1}\sum_{ws \in \mathcal{S}(z)}{p^{-1}_{ws}}(z)$ is the Horvitz--Thompson estimator of constant 1.
As pointed out by \cite{zhao2021reconciling}, the Hajek estimator is a ratio estimator for $\bar{Y}(z) = \bar{Y}(z) / 1$ with the numerator and denominator estimated by their Horvitz--Thompson estimators, respectively. Let $\hat Y_{\haj}$ be the vectorization of $\{ \hat Y_{\haj} (z) \}_{z \in \mathcal{T}}$ in lexicographical order of $z$. We call $\tauhaj = G \hat Y_{\haj}$ the Hajek estimator of $\tau$.

We adopt the design-based framework, which conditions on the potential outcomes and evaluates the sampling properties of $\tauht$ and $\tauhaj$ over the joint distribution of $Z_{ws}$'s. 
Let $\bar{Y}_w(z) = M_w^{-1}\sums Y_{ws}(z)$ be the average potential outcome in whole plot $w$. The covariances of $\tauht$ and $\tauhaj$ under split-plot randomization depend on the {\it scaled between-} and {\it within-whole-plot covariances} of $\{Y_{ws}(z): ws\in\ms; \, z\in\mt\}$ defined as follows: $S_{\HT} = (S_{\HT}(z,z'))_{z,z'\in\mt}$, $S_{\haj}= (S_{\haj}(z,z'))_{z,z'\in\mt}$, and $S_w = (S_w(z,z'))_{z,z'\in\mt}$ for $w = 1, \ldots, W$, where
\begin{eqnarray*}
S_{\HT}(z,z^\prime) &=& (W-1)^{-1}\sum_{w=1}^W\left\{\alpha_w \bar Y_w(z) - \bar Y(z) \right\} \left\{\alpha_w \bar Y_w(z^\prime) - \bar Y(z^\prime)\right\}, \\
S_{\haj}(z,z^\prime) &=& (W-1)^{-1}\sum_{w=1}^W\alpha_w^2\left\{ \bar Y_w(z) - \bar Y(z) \right\} \left\{\bar Y_w(z^\prime) - \bar Y(z^\prime)\right\},\\
S_w(z,z^\prime) &=& (M_w-1)^{-1}\sum_{s=1}^{M_w}\alpha_w^2\left\{  Y_{ws}(z) - \bar Y_w(z) \right\} \left\{ Y_{ws}(z^\prime) - \bar Y_w(z^\prime)\right\}
\end{eqnarray*} 
for $z, z^\prime \in \mathcal{T}$ \citep{mukerjee2019causal, zhao2021reconciling}. 

Let $H=\text{diag}(p_0^{-1},p_1^{-1})\otimes1_{2\times 2} - 1_{4\times 4}$, $H_w=\text{diag}(p_0^{-1},p_1^{-1})\otimes \{\text{diag}(q_{w0}^{-1},q_{w1}^{-1}) - 1_{2\times 2} \}$, and $\Psi =W^{-1}\sum_{w=1}^WM_w^{-1}(H_w\circ S_w)$.
Let $\overline{\alpha^k} = W^{-1} \sumw \alphaw^k$ be the $k$th moment of $(\alphaw)_{w=1}^{W}$ for $k=1,2,4$, and let 
$\overline{Y_{w.}^4(z)} = M_w^{-1} \sums Y_{ws}^4(z)$. 
Condition \ref{cond::1} below  was proposed by \citet{zhao2021reconciling} and gives the regularity conditions for {\it finite population asymptotics} under split-plot randomization \citep{DingCLT}.

\begin{condition}\label{cond::1}
As $W$ goes to infinity, for $a,b=0,1$ and $z \in \mathcal{T}$,
\begin{itemize}
    \item[(i)] $\overline{\alpha^2}=O(1); \ \overline{\alpha^4}=o(W)$;
    \item[(ii)] $p_a$ has a limit in $(0,1)$; for all $w=1,\ldots,W$, $q_{wb}\in [c,1-c]$ for a constant $c \in (0,1/2]$ independent of $W$;
    \item[(iii)] for $*= \HT,\haj$, $S_{*}$, $\bar Y$, and  $\Psi$  have finite limits; 
    \item[(iv)]  $ W^{-1} \max_{w=1, \ldots, W} |\alpha_w \bar{Y}_{w}(z)-\bar{Y}(z)|^2=o(1)$;
       \item[(v)] $W^{-1} \sum_{w=1}^W\alpha_w^2\overline{Y_{w.}^4(z)}=O(1)$; $W^{-2} \sum_{w=1}^W\alpha_w^4\overline{Y_{w.}^4(z)}=o(1)$.
\end{itemize}
\end{condition}

Condition \ref{cond::1}(ii)--(iii) ensure that $\Sigma_{*, \tau\tau} = G ( H\circ S_{*}+\Psi) G^\T$ has a finite 
limit for $* = \HT, \haj$. We will use the same notation to also denote their respective limiting values when no confusion would arise.
Lemma \ref{lem:1} below follows from \cite{zhao2021reconciling} and ensures the consistency and asymptotic normality of $\hat \tau_{*} \ (* = \HT, \haj)$ for estimating $\tau$. 
\begin{lemma}\label{lem:1}
Under Condition \ref{cond::1}, $\sqrt{W} ( \hat \tau_{*} - \tau ) \rs  \mathcal{N}(0,\Sigma_{*, \tau \tau}) $ for $* = \HT, \haj$.
\end{lemma}

\section{Rerandomization in split-plot designs}

\subsection{Rerandomization schemes}
In split-plot designs, we often collect baseline covariates before the experiments, denoted by $x_{ws} = (x_{ws,1},\ldots,x_{ws,L})^\T \in \mathbb{R}^{L}$. 
The cluster randomization in stage (I) and stratified randomization in stage (II) ensure that the covariates are balanced on average at both the whole-plot and subplot levels. However, covariate imbalance often exists in finite samples and subjects subsequent inference to possibly large variability and conditional bias. Rerandomization provides a way to balance covariates in the design stage \citep[see, e.g.,][]{Morgan2012,Morgan2015,Li2018,Li2020factorial,Wang2021,zdrep}.
\citet{Morgan2012} suggested a rerandomization scheme using the Mahalanobis distance of the covariate means under different treatment arms to measure the covariate imbalance in a completely randomized treatment-control experiment. This motivates two rerandomization schemes under split-plot randomization.

Specifically, define
$$
 \hat{x}_{\HT} (z) = N^{-1}\sum_{ws \in \mathcal{S}(z)}{p^{-1}_{ws}}(z)x_{ws},  \quad  \hat {x}_{\haj} (z) = \frac{\hat {x}_{\HT} (z)}{\hat {1}_{\HT} (z)},  
$$
as the Horvitz--Thompson and Hajek estimators of $\bar x = N^{-1} \sumws x_{ws}$ based on units under treatment $z$. Let $\hat x_{*} = ( \hat x_{*}(00), \hat x_{*}(01), \hat x_{*}(10), \hat x_{*}(11) )^\T \in \mathbb{R}^{4 \times L}$ for $* = \HT, \haj$. 
For a $4 \times 1$ contrast vector $g = (g_{00}, g_{01}, g_{10}, g_{11})^\T$, the contrast of $\hat x^\T_*(z)$'s,
$$
    g_{00}\hat x^\T_*(00) + g_{01}\hat x^\T_*(01) + g_{10}\hat x^\T_*(10) + g_{11}\hat x^\T_*(11) = g^\T \hat x_* \in \mathbb{R}^{1 \times L},
$$
provides an intuitive measure of covariate balance under split-plot design. A balanced allocation intuitively has homogeneous $\hat x^\T_*(z)$'s such that $g^\T \hat x_*$ is close to $0_{1 \times L}$.
The contrasts that correspond to $g_{\textsc{a}}$, $g_{\textsc{b}}$, and $g_{\textsc{ab}}$ are 
\begin{eqnarray*}
    g^\T_{\textsc{a}} \hat x_* &=& 2^{-1}\{ \hat x_{*}^\T(10) + \hat x_{*}^\T(11) \} - 2^{-1}\{ \hat x_{*}^\T(00) + \hat x_{*}^\T(01) \} ,\\
   g^\T_{\textsc{b}} \hat x_* &=& 2^{-1}\{ \hat x_{*}^\T(01) + \hat x_{*}^\T(11) \} - 2^{-1}\{ \hat x_{*}^\T(00) + \hat x_{*}^\T(10) \},\\
    g^\T_{\textsc{ab}} \hat x_* &=& \{ \hat x_{*}^\T(00) + \hat x_{*}^\T(11) \} - \{ \hat x_{*}^\T(01) + \hat x_{*}^\T(10) \},
\end{eqnarray*}
respectively. Let $$ \hat \tau_{*,x} = ( g^\T_{\textsc{a}} \hat x_*, g^\T_{\textsc{b}} \hat x_*, g^\T_{\textsc{ab}} \hat x_*)^\T \in \mathbb{R}^{3L}$$
be their concatenation for $* = \HT, \haj$, which is intuitively close to $0_{3L}$ if the allocation is balanced.
We consider two rerandomization schemes based on the Mahalanobis distance between $\hat\tau_{*,x}$ and $0_{3L}$ 
under split-plot randomization.

The first scheme is based on the Mahalanobis distance between $\tauxht$ and $0_{3L}$ 
under split-plot randomization: $\textsc{M}_{\HT}=\tauxht^\T \cov( \tauxht )^{-1}  \tauxht$. For a predetermined threshold $d > 0$,
rerandomization  accepts the treatment assignment if and only if {the following event happens:
$$
\mathcal{M}_{\HT} = \{ \textsc{M}_{\HT} \leq d \}.
$$
The second scheme is based on the Mahalanobis distance  between $\tauxhaj$ and $0_{3L}$ 
under split-plot randomization: $\textsc{M}_{\haj}=\tauxhaj^\T \cova( \tauxhaj )^{-1}  \tauxhaj$, and accepts the treatment assignment if and only if the following event happens:
$$
\mathcal{M}_{\haj} = \{ \textsc{M}_{\haj}\leq d \}.
$$
We define $\textsc{M}_{\haj}$ using the asymptotic covariance $\cova( \tauxhaj )$ due to the complicated form of the exact covariance $\cov(\tauxhaj)$; see Theorem~\ref{thm::clt} in Section~\ref{sec::asy} for more details.

Two treatment effect estimators, $\tauht$ and $\tauhaj$, and two rerandomization schemes, $\mathcal{M}_{\HT}$ and $\mathcal{M}_{\haj}$, give rise to four inferential strategies as their combinations. Nevertheless, it is more natural to consider design and analysis of the same type. Therefore, 
{we will consider $\mathcal{M}_{\HT}$ for rerandomization  if using $\tauht$ for treatment effect estimation, and consider $\mathcal{M}_{\haj}$ if using $\tauhaj$.} To avoid confusion, we will henceforth use {\it classic split-plot randomization} to refer to the standard split-plot randomization without rerandomization.

\subsection{Asymptotic distribution}\label{sec::asy}
For $*=\HT,\haj$, the asymptotic distribution of $\hat \tau_* $ under rerandomization scheme $\mathcal{M}_{*} $ is essentially the conditional asymptotic distribution of $\hat \tau_* $ under classic split-plot randomization given $\mathcal{M}_{*} $, denoted by $\hat \tau_* \mid \mathcal{M}_{*} $ \citep{Li2018}. 
To study them, we start with the unconditional joint asymptotic distributions of  $(\hat\tau^\T_*, \hat \tau^\T_{*,x})^\T$ under classic split-plot randomization.

Let $S_{\HT,xx}$, $S_{\haj,xx}$, $S_{w,xx}$, $S_{\HT,xY(z)}$, $S_{\haj,xY(z)}$, and $S_{w,xY(z)}$ be the scaled between and within whole-plot covariances of $(x_{ws})_{ws \in \mathcal{S}}$ with itself and with $\{Y_{ws}(z)\}_{ws \in \mathcal{S}}$, respectively, analogous to $S_{\HT}(z,z')$, $S_\haj(z,z')$, and $S_{w}(z,z')$.
To avoid too many formulas in the main paper, we relegate their explicit forms 
to the supplementary materials. 
Define
$$
    \Psi_{xx}=W^{-1}\sumw M_w^{-1} (H_w\otimes S_{w,xx}), \quad
    \Psi_{ xY}=W^{-1}\sumw M_w^{-1} (H_w\otimes 1_L)\circ(1_4\otimes S_{w,xY}),
$$
where $S_{w,xY} = ( S_{w,xY(00)}, S_{w,xY(01)}, S_{w,xY(10)}, S_{w,xY(11)}) \in \mathbb{R}^{L \times 4}$. For $* = \HT, \haj$, let
\begin{eqnarray*}
S_{*, xY} &=& ( S_{*,xY(00)}, S_{*,xY(01)}, S_{*,xY(10)}, S_{*,xY(11)}) \in \mathbb{R}^{L \times 4}, \\
\Sigma_{*, xx} & =  &  ( G \otimes I_L ) (H\otimes S_{*,xx}+\Psi_{xx})  ( G \otimes I_L)^\T,\\
\Sigma_{*, x\tau} &=& \Sigma_{*, \tau x}^\T = ( G \otimes I_L) \{ (H\otimes 1_L)\circ(1_4\otimes S_{*,xY}) + \Psi_{xY} \} G^\T.
\end{eqnarray*}
We require Condition~\ref{cond::1} and Condition~\ref{cond::2} below for deriving the joint asymptotic distribution of  $(\hat\tau^\T_*, \hat \tau^\T_{*,x})^\T$ for $* = \HT, \haj$. 
Let $\bar{x}_w = M_w^{-1}\sums x_{ws}$ and $\overline{ ||  x_{w.} ||_\infty^4 } = M_w^{-1} \sums || x_{ws} ||_\infty^4$. 

\begin{condition}\label{cond::2}
As $W$ goes to infinity, 
\begin{itemize}
    \item[(i)] for $*= \HT,\haj$,  $S_{*,xx}, S_{*,xY}$,  $\Psi_{xx}$, and $\Psi_{xY}$  have finite limits; the limits of $S_{*,xx}$ and $\Sigma_{*, xx}$ are invertible;
    \item[(ii)] $ W^{-1} \max_{w=1, \ldots, W} || \alpha_w \bar{x}_{w}-\bar{x}||_\infty^2=o(1)$;  
    \item[(iii)] $W^{-1} \sum_{w=1}^W\alpha_w^2\overline{|| x_{w.} ||_\infty^4}=O(1)$; $W^{-2} \sum_{w=1}^W\alpha_w^4\overline{|| x_{w.} ||_\infty^4}=o(1)$.
\end{itemize}
\end{condition}

Condition~\ref{cond::2} gives the analog of Condition~\ref{cond::1} for the covariates $x_{ws}$'s.
Condition~\ref{cond::1}(ii) and Condition~\ref{cond::2}(i) together ensure that $\Sigma_{*, xx}$, $\Sigma_{*, x \tau}$, and $\Sigma_{*, \tau x}$ all have finite limits for $* = \HT, \haj$. Again, we will use the same notation to also denote their respective limiting values when no confusion would arise.

\begin{theorem}\label{thm::clt}
Under Conditions~\ref{cond::1} and \ref{cond::2}, for $* = \HT, \haj$, 
\begin{eqnarray}
   \sqrt{W} \left( \begin{array}{cc}
        \hat\tau_* - \tau  \\
        \hat \tau_{*,x}
   \end{array} \right) \rs  
   \mathcal{N} \left( 0, \Sigma_* \right), \quad \Sigma_* = \left(\begin{array}{cc}
        \Sigma_{*, \tau\tau} & \Sigma_{*, \tau x}\\  
        \Sigma_{*, x\tau} & \Sigma_{*, xx}\\
   \end{array} \right). \nonumber
\end{eqnarray}
\end{theorem}


Theorem~\ref{thm::clt} ensures the asymptotic joint normality of  $ \hat \tau_{*}$ and $\hat \tau_{*,x}$, and provides the basis for deriving the conditional asymptotic distribution of $\hat\tau_*$ given $\mathcal{M}_{*}$.  
By Theorem~\ref{thm::clt}, the Mahalanobis distance  
$
\textsc{M}_{*}= (\sqrt{W} \hat \tau_{*,x})^\T \Sigma_{*, xx}^{-1}   (\sqrt{W} \hat \tau_{*,x})  
$
converges in distribution to $\chi_{3L}^2$ for both $* = \HT, \haj$. Thus, we can choose the threshold $d$
as the $\alpha$th quantile of $\chi_{3L}^2$ to ensure an asymptotic acceptance rate of $\alpha$ for the rerandomization. 

By Theorem \ref{thm::clt}, the linear projection of $\sqrt{W} \hat \tau_{*}$ onto $ \hat \tau_{*,x}$ equals 
$
\proj(\sqrt{W} \hts \mid  \hat\tau_{*, x}) = \sqrt{W} \tau + \sqrt{W} \Sigma_{*, \tau x} \Sigma^{-1}_{*, xx}\hat\tau_{*, x}$ asymptotically. 
Let $\Sigma_{*, \tau\tau}^{||} = \Sigma_{*, \tau x}\Sigma_{*,xx}^{-1}\Sigma_{*,x\tau}$ denote the asymptotic covariance of $ \proj( \sqrt{W}  \hts \mid \hat\tau_{*, x})$, and let $\Sigma_{*, \tau\tau}^{\bot}=\Sigma_{*, \tau\tau} - \Sigma_{*, \tau\tau}^{||}$ denote that of the residual 
$\res(\sqrt{W}  \hts \mid \hat\tau_{*, x}) = \sqrt{W} \hts - \proj(\sqrt{W} \hts \mid \hat\tau_{*, x})$. 


\begin{theorem}\label{thm::rerand}
Under Conditions~\ref{cond::1} and \ref{cond::2},  for $* = \HT, \haj$, 
$$
\sqrt{W}(\hat\tau_{*}- \tau) \mid \mathcal{M}_{*} \rs  (\Sigma_{*,\tau\tau}^{\bot})^{1/2}\epsilon
            +\Sigma_{*,\tau x}\Sigma_{*,xx}^{-1/2}\zeta_{3L, d},
$$
where $\epsilon \sim \mathcal{N}(0, I_{3})$ is a 3-dimensional standard normal random vector, $\zeta_{3L, d}\sim D \mid D^\T D \leq d$ is a $3L$-dimensional truncated normal random vector with $D \sim \mathcal{N}(0,I_{3L})$, 
and $\epsilon$ and $\zeta_{3L, d}$ are independent.
\end{theorem}

Theorem~\ref{thm::rerand} indicates that the asymptotic distribution of $\hat \tau_{*}$ under rerandomization scheme $\mathcal{M}_{*}$ is the convolution of a normal component and a truncated normal component.
Observe that 
\begina
\sqrt{W} (\hts - \tau) 
&=& \proj(\sqrt{W} \hts \mid \hat\tau_{*, x}) + \res( \sqrt{W} \hts \mid \hat\tau_{*, x}) - \sqrt{W} \tau \\
&=& \sqrt{W} \Sigma_{*, \tau x} \Sigma^{-1}_{*, xx}\hat\tau_{*, x} + \res(\sqrt{W}\hts \mid \hat\tau_{*, x}).
\enda
The term $\res(\sqrt{W}\hts \mid \hat\tau_{*, x})$ is asymptotically independent of $\hat\tau_{*, x}$ under split-plot randomization, and corresponds to the normal vector $(\Sigma_{*,\tau\tau}^{\bot})^{1/2}\epsilon$ unaffected by the rerandomization. 
The term $\sqrt{W} \Sigma_{*, \tau x} \Sigma^{-1}_{*,xx}\hat\tau_{*, x}$ is affected by the rerandomization and corresponds to the truncated normal vector $\Sigma_{*,\tau x}\Sigma_{*,xx}^{-1/2}\zeta_{3L, d}$. 
It extends the asymptotic theory of rerandomization with treatments assigned at only the unit level \citep{Li2018,Li2020factorial,Wang2021} or group level   \citep{Lu2022} to the split-plot designs. 
Moreover, the asymptotic distributions in Theorem~\ref{thm::rerand} are central convex unimodal \citep[][Definition 2 and Proposition 2]{Li2020factorial}.

We use the following notion of peakedness \citep{sherman} to quantify the relative efficiency between different estimators \citep{Li2020factorial, zdrep}.

\begin{definition} 
For two symmetric $m$-dimensional random vectors $U_1$ and $U_2$, we say that $U_1$ is more peaked than $U_2$ if  $\pr (U_1 \in \mathcal{K}) \geq \pr (U_2 \in \mathcal{K} )$ for every symmetric convex set $\mathcal{K} \subset \mathbb{R}^m$.
\end{definition}

Peakedness implies not only smaller covariance, but also  narrower central quantile regions. 
It hence provides a more refined measure than covariance for comparing relative efficiency between estimators with nonnormal asymptotic distributions.
For $* = \HT, \haj$, we say that rerandomization improves the asymptotic efficiency of $\hts $ if the asymptotic distribution of $\hts  - \tau$ under rerandomization, namely $\hts -\tau \mid \mathcal M_*$, is more peaked than that of  $\hts  - \tau$ without rerandomization. Corollary~\ref{cor::varcom} below shows the improvement of asymptotic efficiency of $\hts $ by split-plot rerandomization.


\begin{corollary}\label{cor::varcom}
Under Conditions~\ref{cond::1} and \ref{cond::2}, for $*=\HT,\haj$,
rerandomization by $\mathcal{M}_{*}$ improves the asymptotic efficiency of $\hts$ 
with 
\begin{eqnarray*}
    W \left[\cova( \hat \tau_{*}) - \cova( \hat \tau_{*} \mid \mathcal{M}_{*}) \right] = ( 1 - r_{3L, d}) \Sigma_{*, \tau\tau}^{||} \geq 0,
\end{eqnarray*}
where $r_{3L, d} = \pr( \chi^2_{3L + 2} \leq d )/ \pr( \chi^2_{3L} \leq d ) \leq 1$.
\end{corollary}

\subsection{Estimation of the asymptotic distribution}

By Theorem~\ref{thm::rerand}, to infer $\tau$ based on $\hat \tau_{*}$ under rerandomization scheme $\mathcal{M}_{*}$, we need to estimate $\Sigma_{*, \tau\tau}^{\bot}$ and $\Sigma_{*, x\tau}$ 
for $*=\HT,\haj$. By definition, it suffices to estimate $\Sigma_{*,\tau \tau}$ and $\Sigma_{*, x \tau}$.


Let $\hat Y_w(z) = M_{wb}^{-1} \sum_{s:Z_{ws} = z} Y_{ws}$ be the whole-plot sample mean under treatment $z=(ab)$, and let $A_{w}$ be the level of factor A received by whole plot $w$. 
Define
\begina
\hat S_{\HT}(z,z^\prime) & = & (W_a-1)^{-1}\sum_{w: A_w=a}\left\{\alpha_w \hat Y_w(z) - \hat Y_{\HT}(z) \right\} \left\{\alpha_w \hat Y_w(z^\prime) - \hat Y_{\HT}(z^\prime)\right\},   \\
\hat S_{\haj}(z,z^\prime) & = & (W_a-1)^{-1}\sum_{w: A_w=a} \alphaw^2 \left\{ \hat Y_w(z) - \hat Y_{\haj}(z) \right\} \left\{ \hat Y_w(z^\prime) - \hat Y_{\haj}(z^\prime)\right\}  
\enda
as the sample analogs of $S_{\HT}(z,z^\prime)$ and $S_{\haj}(z,z^\prime)$ for $z=(ab)$ and $z^\prime=(ab^\prime)$ with the same level of factor A. 
For $* = \HT, \haj$, 
\citet[][Theorem 4.2]{zhao2021reconciling} ensures that
$$
\hat\Sigma_{*, \tau\tau}=G\left(                 
            \begin{array}{cc}   
            p_0^{-1}\left(                 
            \begin{array}{cc}   
            \hat S_{*}(00,00) &  \hat S_{*}(00,01) \\  
            \hat S_{*}(00,01) &  \hat S_{*}(01,01) \\ 
            \end{array}\right)
            & 0_{2\times 2}\\  
            0_{2\times 2} & p_1^{-1}\left(                 
            \begin{array}{cc}   
            \hat S_{*}(10,10) &  \hat S_{*}(10,11) \\  
            \hat  S_{*}(10,11) & \hat  S_{*}(11,11) \\ 
            \end{array}\right)\\  
            \end{array}\right)G^\T 
$$
gives an asymptotically conservative estimator of $\Sigma_{*, \tau\tau}$ under classic split-plot randomization.

Let $\hat Y_{\HT,ws}(z) =  \mathcal{I}(Z_{ws}=z) p_{ws}(z)^{-1}  Y_{ws}$ and $\hat Y_{\HT,w}(z) = M_{w}^{-1} \sums \mathcal{I}(Z_{ws}=z) p_{ws}(z)^{-1}  Y_{ws}$ 
be the Horvitz--Thompson estimators of $Y_{ws}(z)$ and $\bar Y_w(z)$, respectively. 
Let $\hat S_{\HT,xY}$, $\hat S_{\haj,xY}$, and $\hat S_{w,xY}$ be the sample analogs of $S_{\HT,xY}$, $S_{\haj,xY}$, and $S_{w,xY}$, respectively, with $Y_{ws}(z)$, $\bar Y_w(z)$, and $\bar Y(z)$ estimated by $\hat Y_{\HT,ws}(z)$, $\hat Y_{\HT,w}(z)$, and $\hat Y_{\HT}(z)$, respectively. 
We can then estimate $\Sigma_{*,  x\tau}$ by 
\begina
\hat\Sigma_{*,  x\tau}  = (G\otimes I_L)\left\{(H\otimes 1_L)\circ(1_4\otimes \hat S_{*, xY}) + \hat\Psi_{ xY}\right\}G^\T,
\enda
where $\hat\Psi_{xY} = W^{-1}\sumw M_w^{-1}(H_w\circ \hat S_{w, xY})$.
This yields
$$
 \hat\Sigma_{*} = \left(                 
            \begin{array}{cc}   
            \hat\Sigma_{*, \tau\tau} & \hat\Sigma_{*,  \tau x}\\  
            \hat\Sigma_{*,  x \tau} & \Sigma_{*, xx}\\ 
            \end{array}\right), 
$$
where $\hat\Sigma_{*,  \tau x}  = \hat\Sigma^\T_{*,  x \tau}$, as a plug-in estimator of $\Sigma_{*} \ (*  = \HT, \haj)$.

\begin{theorem}\label{thm::varest}
Under Conditions~\ref{cond::1} and \ref{cond::2}, 
for $* = \HT, \haj$, 
    \[
        \begin{split}
            ( \hat \Sigma_{*} - \Sigma_{*} )  \mid \mathcal{M}_{*} = \left(                 
            \begin{array}{cc}   
            GS_{*}G^\T & 0_{3 \times 3L}\\  
            0_{3L \times 3} & 0_{3L \times 3L}\\ 
            \end{array}\right)
            + o_{\mathbb P}(1). 
        \end{split}
    \]
\end{theorem}

As $GS_{*}G^\T$ is positive semi-definite, Theorem~\ref{thm::varest} shows that $\hat \Sigma_{*, \tau\tau}$ is an asymptotically conservative estimator of $\Sigma_{*, \tau\tau}$ and $\hat\Sigma_{*, x\tau}$ is a consistent estimator of $\Sigma_{*, x\tau}$ under split-plot rerandomization. Thus, $\hat \Sigma_{*, \tau\tau}^{||} = {\hat \Sigma_{*,\tau x}} \Sigma_{*,xx}^{-1} \hat \Sigma_{*,x\tau} $ is a consistent estimator of $\Sigma_{*, \tau\tau}^{||} $ and $\hat \Sigma_{*, \tau\tau}^{\bot} = \hat \Sigma_{*, \tau\tau} - \hat \Sigma_{*, \tau\tau}^{||}$ is a conservative estimator of $\Sigma_{*, \tau\tau}^{\bot}$. Therefore, the asymptotic distribution of $\sqrt{W}( \hat \tau_{*} - \tau ) \mid \mathcal{M}_{*}$ can be conservatively estimated by
$
 \phi_* = (\hat \Sigma_{*,\tau\tau}^{\bot})^{1/2}\epsilon
            + \hat \Sigma_{*,\tau x}\Sigma_{*,xx}^{-1/2}\zeta_{3L, d}.
$
Suppose that the limit of $\Sigma_{*, \tau\tau}^{\bot}$ is invertable, then $\hat \Sigma_{*,\tau\tau}^{\bot}$ is invertable with probability tending to one. Let $\hat c_{*, 1 - \xi}$ and $\chi^2_{3, 1 - \xi}$  ($0 < \xi < 1$) be the $1 - \xi$ quantiles of $ \phi_*^\T (\hat \Sigma_{*,\tau\tau}^{\bot})^{-1}  \phi_*$ and $\chi^2_{3}$, 
respectively. Corollary~\ref{cor::CI} below provides asymptotically conservative confidence regions for $\tau$ and demonstrates that rerandomization generally improves the inference efficiency.

\begin{corollary}\label{cor::CI}
Suppose that the limit of $\Sigma_{*, \tau\tau}^{\bot}$ is invertable. Under Conditions~\ref{cond::1} and \ref{cond::2}, for $* = \HT, \haj$, the  Wald-type confidence region
$
\{ \tau:  W (\hat \tau_{*} - \tau)^\T (\hat \Sigma_{*,\tau\tau}^{\bot})^{-1} (\hat \tau_{*} - \tau) \leq \hat c_{*, 1 - \xi}  \}
$
has asymptotic coverage rate greater than or equal to $1 - \xi$ under the corresponding  split-plot rerandomization scheme. Moreover, the area of the above confidence region is smaller than or equal to that of the confidence region 
$
\{ \tau:  W (\hat \tau_{*} - \tau)^\T \hat \Sigma_{*,\tau\tau}^{-1} (\hat \tau_{*} - \tau) \leq \chi^2_{3, 1 - \xi}  \}
$
under the classic split-plot randomization.
\end{corollary}






\section{Covariate Adjustment under Rerandomization}

The discussion so far concerned rerandomization that enforces covariate balance in the design stage. 
Alternatively, we can adjust for covariate imbalance in the analysis stage.
\cite{Li2020} and \cite{Wang2021} showed the duality of rerandomization and regression adjustment for improving efficiency in completely randomized and stratified treatment-control experiments, respectively.
In this section, we extend the discussion to the method and design-based theory of the combination of rerandomization and covariate adjustment in $2^2$ split-plot designs. 
We consider two strategies for covariate adjustment 
 for each of the Horvitz--Thompson and Hajek estimators, and derive their design-based properties under split-plot rerandomization. 
The first strategy follows the regression formulation by \cite{zhao2021reconciling}.
The second strategy is new and approaches covariate adjustment from a projection or conditional inference perspective.

Let $v_{ws} \in \mathbb{R}^{J}$ denote the covariates used in the analysis stage.
We allow the analysis stage to use more covariates than the design stage 
in the sense that $ x_{ws} = C v_{ws} $ for some matrix $C \in \mathbb R^{L \times J}$ ($J \geq L$). 
Let $\bar v = N^{-1} \sum_{ws \in \mathcal S} v_{ws}$, $\bar v_{w} = M_w^{-1} \sum_{s=1}^{M_w} v_{ws}$, and $\hat v_w (z) = M_{wb}^{-1} \sum_{s: Z_{ws}=z} v_{ws}$ 
for $w=1,\ldots,W$ and $z = (ab) \in \mt$. 
For $*=\HT,\haj$, define  $S_{*,vv}$, $S_{w,vv}$, $\Psi_{vv}$, $S_{*,vY(z)}$, $S_{w,vY(z)}$, $S_{*,vY}$, $S_{w,vY}$, $\Psi_{ vY}$, $\Sigma_{*, vv}$, $\Sigma_{*,v \tau}$, $\Sigma_{*, \tau v}$, $\hat {v}_{*}(z)$, $\hat {v}_{*}$, and $\hat \tau_{*, v}$ similarly to $S_{*,xx}$, $S_{w,xx}$, $\Psi_{xx}$, $S_{*,xY(z)}$, $S_{w,xY(z)}$, $S_{*,xY}$, $S_{w,xY}$, $\Psi_{xY}$, $\Sigma_{*,xx}$, $\Sigma_{*,x \tau}$, $\Sigma_{*, \tau x}$, $\hat {x}_{*}(z)$, $\hat {x}_{*}$, and $\hat \tau_{*, x}$, with $x_{ws}$ replaced by $v_{ws}$. 


\subsection{Regression with treatment-covariate interactions}

Regression adjustment provides a convenient 
way to adjust for covariate imbalance in the analysis stage. 
For observed data $\{(y_i, u_i):  i \in \mathcal J, \ y_i \in \mathbb R, \ u_i \in \mathbb R^m\}$, where $\mathcal J$ denotes the index set, denote by $y_i \sim u_i$  the linear regression of $y_i $ on $u_i$ over $i \in \mathcal J$. 
\cite{zhao2021reconciling} showed that the Horvitz--Thompson and Hajek estimators $\hat\tau_* \ (* = \HT, \haj)$ can be recovered from the ordinary least squares (ols) fit of the aggregate regression
\beginy\label{eq:ag}
        \alphaw \hat Y_w(A_w b) \sim \mathcal{I}(A_w b =00)+ \mathcal{I}(A_w b=01)+ \mathcal{I}(A_w b=10)+ \mathcal{I}(A_w b=11)
\endy
over $\{(w,b): w = 1,\ldots, W; \ b = 0,1\}$ and the weighted least squares (wls) fit of 
\beginy\label{eq:wls}
        Y_{ws} \sim \mathcal{I}(Z_{ws}=00)+ \mathcal{I}(Z_{ws}=01)+ \mathcal{I}(Z_{ws}=10)+ \mathcal{I}(Z_{ws}=11)
\endy
over $ws \in \mathcal S$, respectively, and recommended including full interactions between the treatment indicators and centered covariates for regression adjustment. 
In particular, let $\tba $ and $\tva$ 
be the ols coefficient vector and associated cluster-robust covariance from \eqref{eq:ag}, where we use 
the subscript ``ag" to signify the use of whole-plot aggregate outcomes and covariates in forming the regression \citep{Abadie2008,Basse2018, imai2020, Su2021}.  
Let $\tbw$ and $\tvw $ be the wls coefficient vector and associated cluster-robust covariance from \eqref{eq:wls}, where we weight subplot $ws$ by the inverse of its realized inclusion probability $p_{ws} (Z_{ws})$. 
\cite{zhao2021reconciling} showed that
$\tba  = \hat{Y}_{\HT}$ and $\tbw = \hat{Y}_{\haj}$, with $\tva $ and $\tvw $ being asymptotically conservative for estimating the true sampling covariances. This justifies the large-sample Wald-type inference of $\tau$ based on $(G\tba, G\tva G^\T)$ and $(G\tbw, G\tvw G^\T)$.
Further let 
\beginy\label{eq:ag_v}
 \alphaw \hat Y_w(A_w b) & \sim & \sum_{z \in \mathcal{T} } \mathcal{I}(A_w b =z) + \sum_{z \in \mathcal{T} } \mathcal{I}(A_w b =z)\alphaw \{\hat v_w(A_w b) - \bar v\}, \\
 Y_{ws} & \sim &  \sum_{z \in \mathcal{T} } \mathcal{I}(Z_{ws} =z) + \sum_{z \in \mathcal{T} } \mathcal{I}(Z_{ws} =z) ( v_{ws} - \bar v) \label{eq:wls_v}
\endy
be the fully interacted variants of \eqref{eq:ag} and \eqref{eq:wls}. 
Let $\tbal $ and $\tbwl $ denote the ols and wls coefficient vectors of $\{\mathcal{I}(A_w b =z)\}_{z \in \mathcal{T}}$ and $\{\mathcal{I}(Z_{ws} =z)\}_{z \in \mathcal{T}} $ from \eqref{eq:ag_v} and \eqref{eq:wls_v}, respectively, with $\tval $ and $\tvwl $ as the associated cluster-robust covariances. 
They form the regression-adjusted counterparts of $(\tba , \tva )$ and $(\tbw , \tvw )$. 
We use the subscript ``$\L$" to signify \cite{lin2013}, who proposed the fully interacted adjustment under completely randomized experiments.

Let $\tauhtL = G \tbal  $ and $\tauhajL = G \tbwl $ be the corresponding {\it regression-adjusted Horvitz--Thompson and Hajek estimators}  of $\tau$, with ${ \hat \Sigma}_{\HT,\L, \tau \tau} = W G \tval  G^\T $ and ${ \hat \Sigma}_{\haj,\L, \tau \tau} = W G \tvwl  G^\T $ as the associated cluster-robust covariance estimators up to a factor of $W$. 
\citet[][Theorem 6.2]{zhao2021reconciling} ensured the asymptotic validity of $(\hat\tau_{*,\L},   \hat \Sigma_{*,\L, \tau \tau})$  for inferring $\tau$ under the classic split-plot randomization. 
Theorem~\ref{thm::rerandreg} below extends their results and presents the asymptotic properties of $(\hat\tau_{*,\L},   \hat \Sigma_{*,\L, \tau \tau})$ under split-plot rerandomization.

Let $\hat \gamma_{\ag,z}$ and $\hat \gamma_{\wls,z}$ be the coefficient vectors of $\mathcal{I}(A_w b =z)\alphaw  \{ \hat v_w(A_w b) - \bar v\}$ and $\mathcal{I}(Z_{ws} =z) ( v_{ws} - \bar v)$ from the ols and wls fits of \eqref{eq:ag_v} and \eqref{eq:wls_v}, respectively. 
Under Condition~\ref{cond::3} below, $\hat \gamma_{\ag,z}$ and $\hat \gamma_{\wls,z}$ have finite probability limits, denoted by $\gamma_{\ag,z}$ and $\gamma_{\wls,z}$ respectively, under split-plot rerandomization.
We give the exact formulas of $\gamma_{\ag,z}$ and $\gamma_{\wls,z}$ in the supplementary materials. 
Define covariate-adjusted potential outcomes $Y_{ws}(z; \gamma_{\dagger,z}) = Y_{ws}(z) - (v_{ws} - \bar v)^\T \gamma_{\dagger,z}$ for $\dagger = \ag, \wls$ and $z \in \mathcal{T}$. 
Define 
\begin{eqnarray*}
\Sigma_{*,\L} = \left(\begin{array}{cc}
        \Sigma_{*, \L, \tau\tau} & \Sigma_{*, \L, \tau x}\\  
        \Sigma_{*, \L, x \tau} &  \Sigma_{*, xx} \\
   \end{array} \right) \quad (* = \HT, \haj)
\end{eqnarray*}
similarly to $\Sigma_{*}$ with $Y_{ws}(z)$ replaced by $Y_{ws}(z; \gamma_{\ag,z})$ and $Y_{ws}(z; \gamma_{\wls,z})$, respectively, for $* = \HT$ and $* = \haj$. 
Applying Theorem \ref{thm::clt} to the covariate-adjusted potential outcomes ensures that $\Sigma_{*, \lin}$ gives the asymptotic covariance matrix of $\sqrt W(\hat \tau_{*,\L}^\T, \hat \tau_{*,x}^\T)^\T$.  
The $\hat \Sigma_{*,\lin,\tau\tau}$ from regression thus gives a convenient estimator of  $\Sigma_{*, \L, \tau\tau} = W\cova(\hat \tau_{*,\L})$. 

Let $Q_{vv} = (N-1)^{-1} \sum_{ws \in \mathcal{S}} (v_{ws}-\bar v)(v_{ws}-\bar v)^\T$ and $Q_{vY(z)} = (N-1)^{-1} \sum_{ws \in \mathcal{S}}  (v_{ws} - \bar v)Y_{ws}(z)$ be the finite population covariances of $(v_{ws})_{ws \in \mathcal{S}}$ with itself and $\{Y_{ws}(z)\}_{ws \in \mathcal{S}}$, respectively.

\begin{condition}\label{cond::3} (i) Condition~\ref{cond::2} holds with $x_{ws}$ replaced by $v_{ws}$; (ii) as $M$ goes to infinity, $Q_{vv}$ and $Q_{vY(z)}$
have finite limits, and the limit of $Q_{vv}$ is invertible.
\end{condition}

Conditions~\ref{cond::1}--\ref{cond::3} ensure that $\Sigma_{*, \L, \tau\tau}$, $ \Sigma_{*, \L, \tau x}$,  $\Sigma_{*, \L, x \tau}$, and $ \Sigma_{*,  \L, xx}$ all have finite limits  for $* = \HT, \haj$. We will use the same notation to also denote their respective limiting values when no confusion would arise. Recall that $\hat \Sigma_{*,\lin,\tau\tau}$ gives a convenient estimator of  $\Sigma_{*, \L, \tau\tau} = W\cova(\hat \tau_{*,\L})$. 
Let $\Sigma_{*, \L, \tau\tau}^{||} = \Sigma_{*, \L,\tau x}\Sigma_{*,xx}^{-1}\Sigma_{*,\L,x\tau}$ denote the covariance of the linear projection of $\sqrt{W} \hat \tau_{*,\L}$ onto $\hat \tau_{*,x}$ analogous to $\Sigma_{*, \tau\tau}^{||}$, and let $\Sigma_{*, \L,\tau\tau}^{\bot}=\Sigma_{*, \L,\tau\tau} - \Sigma_{*, \L, \tau\tau}^{||}$ denote the corresponding covariance of the residual. 
Let $\hat \Sigma_{*, \L, \tau x} = \hat \Sigma_{*, \L,  x\tau}^\T$  be the plug-in estimators of $ \Sigma_{*, \L, \tau x}=  \Sigma_{*, \L,  x\tau}^\T$, which are defined similarly to $\hat \Sigma_{*, \tau x} = \hat \Sigma_{*,  x\tau}^\T$ with  $Y_{ws}(z)$ replaced by $Y_{ws}(z; \hat \gamma_{\ag,z})$ and $Y_{ws}(z; \hat \gamma_{\wls,z})$, respectively, for $* = \HT$ and $* = \haj$.
Let $\hat \Sigma_{*, \L,\tau\tau}^{\bot} = \hat\Sigma_{*, \L,\tau\tau} - \hat\Sigma_{*, \L,\tau x}\Sigma_{*,xx}^{-1}\hat\Sigma_{*,\L,x\tau}$ be the corresponding estimator of $\Sigma_{*, \L,\tau\tau}^{\bot}$.

\begin{theorem}\label{thm::rerandreg}
Under Conditions~\ref{cond::1}--\ref{cond::3}, 
for $* = \HT, \haj$, 
\begin{eqnarray*}
 \sqrt{W}(\hat\tau_{*, \L}- \tau) \mid \mathcal{M}_{*} &\rs & (\Sigma_{*, \L, \tau\tau}^{\bot})^{1/2}\epsilon+\Sigma_{*, \L, \tau x}\Sigma_{*,  xx}^{-1/2}\zeta_{3L, d},
\end{eqnarray*}
where $\epsilon \sim \mathcal{N}(0, I_{3})$ is a 3-dimensional standard normal random vector, $\zeta_{3L, d}\sim D \mid D^\T D \leq d$ is a $3L$-dimensional truncated normal random vector with $D \sim \mathcal{N}(0,I_{3L})$, 
and $\epsilon$ and $\zeta_{3L, d}$ are independent. Moreover,
  \[ 
        \begin{split} 
         ( { \hat \Sigma}_{*,\L, \tau \tau} -  \Sigma_{*,\L, \tau \tau} )  \mid \mathcal{M}_{*} &=  GS_{*,\L}G^T +  o_{\mathbb P}(1), \quad ( \hat \Sigma_{*, \L, \tau x} -  \Sigma_{*, \L, \tau x} )  \mid \mathcal{M}_{*} =  o_{\mathbb P}(1), \\
        \hat \Sigma_{*, \L, \tau\tau}^{\bot} - \Sigma_{*, \L, \tau\tau}^{\bot} \mid \mathcal{M}_{*} &=  GS_{*,\L}G^T +  o_{\mathbb P}(1),
        \end{split}
    \]
where $S_{*,\L}$ is a positive semi-definite matrix.
\end{theorem}

Theorem~\ref{thm::rerandreg} implies that 
the cluster-robust covariance estimator $\hat \Sigma_{*,\L, \tau \tau}$ 
is asymptotically conservative for $\Sigma_{*,\L, \tau \tau}$. 
As the truncated normal distribution is more peaked than the normal distribution and ${ \hat \Sigma}_{*,\L, \tau \tau}$ $ \geq \Sigma_{*,\L, \tau \tau} \geq \cova\{ \sqrt{W}(\hat\tau_{*, \L}- \tau) \mid \mathcal{M}_{*}\}$ holds in probability, 
we can still use the normal approximation with the cluster-robust covariance to construct Wald-type confidence regions as
$
\{ \tau:  W (\hat \tau_{*,\L} - \tau)^\T { \hat \Sigma}_{*,\L, \tau\tau}^{-1} (\hat \tau_{*,\L} - \tau) \leq \chi^2_{3, 1 - \xi}  \}$.
Such confidence regions, whereas asymptotically valid, are overconservative.
A less conservative confidence region is 
$
\{ \tau:  W (\hat \tau_{*,\L} - \tau)^\T (\hat \Sigma_{*,\L,\tau\tau}^{\bot})^{-1} (\hat \tau_{*,\L} - \tau) \leq \hat c_{*, \L, 1 - \xi}  \},
$
where $\hat c_{*, \L, 1 - \xi} $ is defined similarly to $\hat c_{*, 1 - \xi}$, i.e.,  the $1 - \xi$ quantile of $ \phi_{*, \L}^\T (\hat \Sigma_{*, \L, \tau\tau}^{\bot})^{-1}  \phi_{*, \L}$ with $\phi_{*, \L} = (\hat \Sigma_{*, \L,  \tau\tau}^{\bot})^{1/2}\epsilon
            + \hat \Sigma_{*, \L, \tau x}\Sigma_{*,xx}^{-1/2}\zeta_{3L, d}$.

Theorem~\ref{thm::rerandreg} extends \citet[][Theorem 1]{Li2018} and \citet[][Theorem 2]{Li2020factorial} 
to rerandomization under split-plot designs. 
Distinct from these previous results, the asymptotic distributions of the regression-adjusted estimators under split-plot rerandomization are generally not normal, but convolutions of a normal component and a truncated normal component. 
The reason is as follows: as shown in the supplementary materials, for $* = \HT, \haj$, the regression adjustments are equivalent to linearly projecting 
$\hat {Y}_{*}(z)$ onto $ \hat {v}_{*}(z)$ for $z \in  \mathcal{T}$ separately; however, the separate projection differs from the joint projection of $\hat Y_{*}$ onto $ \hat {v}_{*}$ due to the dependence structure of $\{ \hat {v}_{*}(z)\}_{z \in \mathcal{T}}$, such that $\Sigma_{*,\L,  \tau x} \neq 0$ in general.
Moreover, the regression-adjusted estimators cannot guarantee efficiency gains over the unadjusted counterparts.

In some special cases, for example, when only whole-plot covariates 
 are used with $v_{ws} = \bar v_{w}$ or more generally, $\Psi_{vv} = o(1)$,
 the truncated normal component can disappear and the regression-adjusted Horvitz--Thompson estimator $\hat\tau_{\HT, \L}$ is asymptotically more efficient than its unadjusted counterpart under split-plot rerandomization.  Corollary~\ref{cor::varl} below shows the asymptotic distribution of $\hat\tau_{\HT, \L} \mid \mathcal{M}_{\HT}$ when $\Psi_{vv} = o(1)$, and ensures its efficiency gain over the unadjusted counterpart.

\begin{corollary}\label{cor::varl}
Under Condition~\ref{cond::1}--\ref{cond::3},
if $\Psi_{vv} = o(1)$, then $\Sigma_{\HT,\L, \tau x} = o(1)$, $\Sigma_{\HT, \L, \tau\tau}^{\bot} = \Sigma_{\HT, \L, \tau\tau} + o(1)$,
\begin{eqnarray*}
 \sqrt{W}(\hat\tau_{\HT, \L}- \tau) \mid \mathcal{M}_{\HT}   \rs    (\Sigma_{\HT, \L, \tau\tau}^{\bot})^{1/2}\epsilon, \quad ( { \hat \Sigma}_{\HT,\L, \tau \tau} -  \Sigma^{\bot}_{\HT,\L, \tau \tau} )  \mid \mathcal{M}_{\HT} = GS_{\HT,\L}G^\T +   o_{\mathbb P}(1).
\end{eqnarray*}
Moreover, $\Sigma_{\HT, \tau\tau}^{\bot} \geq \Sigma_{\HT, \L, \tau\tau}^{\bot}$ and
\begin{eqnarray*}
    W\left[\cova(\hat\tau_{\HT} \mid \mathcal{M}_{\HT}) - \cova(\hat\tau_{\HT, \L} \mid \mathcal{M}_{\HT}) \right] &=& \Sigma_{\HT, \tau\tau}^{\bot} - \Sigma_{\HT, \L, \tau\tau}^{\bot} + r_{3L, d}\Sigma_{\HT, \tau\tau}^{||} \geq 0.
\end{eqnarray*}
\end{corollary}

Two sufficient conditions for $\Psi_{vv} = o(1)$ are (i) $v_{ws} = \bar v_w$ and (ii) $(S_{w,vv})_{w=1}^W$ are uniformly bounded while $M_w$ goes to infinity for all $w$. 
 Corollary~\ref{cor::varl} implies that under either of these two conditions, 
 we can ensure efficiency gain of the Horvitz--Thompson estimator by regression with treatment-covariate interactions under split-plot rerandomization. 

We cannot guarantee efficiency improvement for the regression-adjusted Hajek estimator under the condition $\Psi_{vv} = o(1)$.
However, when the whole-plot total potential outcomes are more heterogeneous than the whole-plot average potential outcomes, 
$\hat \tau_{\haj,\L}$ can be more efficient than $\hat \tau_{\HT,\L}$ under their corresponding rerandomization schemes.



\subsection{Covariate adjustment by removing the conditional bias}

By Theorem~\ref{thm::rerandreg}, the regression-adjusted estimators cannot guarantee efficiency gains when heterogeneous unit-level covariates are used in the analysis stage under split-plot randomization or rerandomization. To address this issue, we propose a new covariate-adjusted estimator based on a projection or conditional inference perspective. 


Applying Theorem~\ref{thm::clt} to $Y_{ws}(z)$ and $v_{ws}$, $\sqrt{W}( (\hat \tau_{*} - \tau)^\T, \hat \tau^\T_{*,v})^\T$ is asymptotically jointly normal. Then conditional on $\hat \tau_{*,v}$, $\sqrt{W}  ( \hat \tau_{*} - \tau )$ is asymptotically normal with mean $ \sqrt{W} \Sigma_{*,\tau v} \Sigma_{*, vv}^{-1}  \hat \tau_{*,v}$ and covariance $\Sigma_{*,\P,\tau \tau}^\bot = \Sigma_{*, \tau \tau} - \Sigma_{*, \tau v} \Sigma_{*, vv}^{-1} \Sigma_{*, v \tau} \leq \Sigma_{*, \tau \tau}$.
Let $\hat \Sigma_{*,\tau v} = \hat \Sigma^\T_{*,v \tau} $ be a consistent estimator of $\Sigma_{*,\tau v} = \Sigma^\T_{*,v \tau}$, defined similarly to $\hat \Sigma_{*,\tau x} = \hat \Sigma^\T_{*,x \tau}$ with $x_{ws}$ replaced by $v_{ws}$.
We define 
\begina
\hat \tau_{*,\P} = \hat \tau_{*} - \hat \Sigma_{*,\tau v}  \Sigma_{*, vv}^{-1}  \hat \tau_{*,v} 
\enda
as a conditionally consistent estimator of $\tau$.
Since $W \cova( \hat \tau_{*,\P} )= \Sigma_{*,\P, \tau \tau}^\bot = W \min_{\Gamma} \cova( \hat \tau_{*} - \Gamma  \hat \tau_{*,v} )$, $\hat \tau_{*,\P}$
is asymptotically equivalent to the linear projection of $ \hat \tau_{*} $ onto $ \hat \tau_{*, v} $, referred to as the {\it projection estimator} of $\tau$. Let $\hat \Sigma_{*,\P, \tau \tau}^\bot = \hat \Sigma_{*, \tau \tau} - \hat \Sigma_{*,\tau v} \Sigma_{*, vv}^{-1} \hat \Sigma_{*, v \tau}$.


\begin{theorem}\label{thm::project}
Under Conditions~\ref{cond::1}--\ref{cond::3}, for $* = \HT, \haj$, 
\begin{eqnarray*}
 \sqrt{W}(\hat\tau_{*, \P}- \tau) \mid \mathcal{M}_{*} &\rs & 
 (\Sigma_{*,\P, \tau \tau}^\bot)^{1/2}\epsilon, \quad ( \hat \Sigma_{*,\P, \tau \tau}^\bot -  \Sigma_{*,\P, \tau \tau}^\bot )  \mid \mathcal{M}_{*} = GS_{*}G^\T +   o_{\mathbb P}(1) .
\end{eqnarray*}
Moreover,
\begin{eqnarray*}
    W\left[\cova(\hat\tau_{*} \mid \mathcal{M}_{*}) - \cova(\hat\tau_{*,\P} \mid \mathcal{M}_{*}) \right] = 
    \Sigma_{*,\tau \tau}^{\bot} - \Sigma_{*,\P,\tau\tau}^{\bot} + r_{3L, d} \Sigma_{*,\tau \tau}^{||} \geq 0.
\end{eqnarray*}
\end{theorem}

Theorem~\ref{thm::project} implies that, under the rerandomization scheme $\mathcal{M}_{*}$, the treatment effect estimator $\hat\tau_{*, \P}$ is consistent and asymptotically normal, and the covariance estimator $\hat \Sigma_{*,\P, \tau \tau}^\bot$ is asymptotically conservative. Moreover, $\hat\tau_{*, \P}$ improves the efficiency of $\hat \tau_{*}$ without requiring $\Psi_{vv} = o(1)$. 
Based on this theorem, an asymptotically conservative Wald-type confidence region for $\tau$ is
$
\{ \tau:  W (\hat \tau_{*,\P} - \tau)^\T (\hat \Sigma_{*,\P, \tau \tau}^\bot)^{-1} (\hat \tau_{*,\P} - \tau) \leq \chi^2_{3, 1 - \xi}  \}. 
$

\subsection{Relative efficiency of different rerandomization and estimation schemes}

We have introduced the regression-adjusted and projection-based variants for both the Horvitz--Thompson and Hajek estimators of the average treatment effects.
Corollary~\ref{cor::comrr} below gives the relative efficiency between the Horvitz--Thompson and Hajek estimators either with or without covariate adjustment under their respective rerandomization schemes.

Let 
$Q_{\w,vv} = (N-1)^{-1} \sumws (v_{ws}-\bar v_w)(v_{ws}-\bar v_w)^\T$ be a variant of $Q_{vv}$ with $v_{ws}$ centered by the whole-plot average $\bar v_w$ instead of $\bar v$.
It is then a weighted average of the $S_{w,vv}$'s with $Q_{\w,vv} = (N-1)^{-1}\sumw (M_w-1) \alpha_w^{-2} S_{w,vv}$. 
Similarly define $Q_{\w,vY(z)}=(N-1)^{-1} \sumws  (v_{ws}-\bar v_w)\left\{ Y_{ws}(z)-\bar Y_w (z)\right\}$ and $Q_{\w} (z,z^\prime) = (N-1)^{-1} \sumws  \left\{ Y_{ws}(z)-\bar Y_w (z)\right\} \left\{ Y_{ws}(z^\prime)-\bar Y_w (z^\prime)\right\}$
for $z, z^\prime \in \mathcal{T}$. We use the subscript ``in" to signify within whole-plot covariances.

\begin{condition}\label{cond::Qcond}
As $W$ goes to infinity, $Q_{\w,vv} = o(1)$ and $Q_{\w} (z,z)  = O(1)$ for all $z \in \mathcal{T}$.
\end{condition}

\begin{remark}
If only whole-plot covariates are used, then $Q_{\w,vv} = 0$ and $\Psi_{vv} =0$. Both $Q_{\w,vv}$ and $\Psi_{vv}$ measure the variability of covariates within whole plots, but $Q_{\w,vv}=o(1)$ is a stricter condition than $\Psi_{vv} = o(1)$. See the supplementary materials for details.



\end{remark}

\begin{corollary}\label{cor::comrr}
Under Conditions~\ref{cond::1}--\ref{cond::3},  
\begin{itemize}
    \item[\textup{(i)}] 
    $\cova(\hat \tau_{\haj} \mid 
    \mathcal{M}_{\haj}) = \cova(\hat \tau_{\HT}  \mid 
    \mathcal{M}_{\HT})$ and $\cova(\hat \tau_{\haj,\P}\mid \mathcal{M}_{\haj}) = \cova(\hat \tau_{\HT,\P}\mid \mathcal{M}_{\HT})$
    if $\bar x = 0$, $\bar Y(z)=0$ for all $z$ or $\alphaw=1$ for all $w$;\\
    $\cova(\hat \tau_{\haj,\L}\mid \mathcal{M}_{\haj}) = \cova(\hat \tau_{\HT,\L}\mid \mathcal{M}_{\HT}) $ if the design is uniform 
    and Condition~\ref{cond::Qcond} holds;
    \item[\textup{(ii)}] Further assume that $\Psi_{vv} = o(1)$, then 
\begina
\cova(\hat \tau_{\haj} \mid 
    \mathcal{M}_{\haj}) \leq \cova(\hat \tau_{\HT}  \mid 
    \mathcal{M}_{\HT}), \quad \cova(\hat \tau_{\haj,\diamond}\mid \mathcal{M}_{\haj}) \leq \cova(\hat \tau_{\HT,\diamond}\mid \mathcal{M}_{\HT})\quad (\diamond= \L, \P)
\enda if $\bar Y_w(z)$ are constant over all $w$, and 
 \begina
 \cova(\hat \tau_{\haj} \mid 
    \mathcal{M}_{\haj}) \geq \cova(\hat \tau_{\HT}  \mid 
    \mathcal{M}_{\HT}), \quad \cova(\hat \tau_{\haj,\diamond}\mid \mathcal{M}_{\haj}) \geq \cova(\hat \tau_{\HT,\diamond}\mid \mathcal{M}_{\HT})\quad (\diamond = \L, \P)
    \enda
     if $\alphaw \bar Y_w(z)$ are constant over all $w$.
\end{itemize}
\end{corollary}

Corollary~\ref{cor::comrr}{(i)} implies that $\hat \tau_{\haj,\P}$ and $\hat \tau_{\HT,\P}$ are asymptotically equally efficient if the whole plots are of equal sizes, and $\hat \tau_{\haj,\L}$ and $\hat \tau_{\HT,\L}$ are asymptotically equally efficient under uniform design and Condition~\ref{cond::Qcond}. 
Suppose that the within whole-plot covariance of covariates is neglectable, i.e., $\Psi_{vv} = o(1)$. Corollary~\ref{cor::comrr}(ii) implies that, under split-plot rerandomization, 
$\hat \tau_{\haj}$, $\hat \tau_{\haj,\P}$, and $\hat \tau_{\haj,\L}$ are asymptotically more efficient than $\hat \tau_{\HT}$, $\hat \tau_{\HT,\P}$ and $\hat \tau_{\HT,\L}$, respectively, if the whole plots have similar average potential outcomes, and 
vice versa if the whole plots have similar total potential outcomes. 
As the whole-plot totals are often more heterogeneous than the whole-plot averages in practice, we prefer the Hajek estimators and the associated rerandomization scheme over the Horvitz--Thompson estimators and the associated rerandomization scheme in general.

Next, we study the relative efficiency of the regression-adjusted estimators versus the projection estimators.
Let $\hat \tau_{\HT,\L,\alpha}$ denote the analog of $\hat \tau_{\HT,\L}$ that further includes the centered whole-plot size factor $\alphaw - 1$  as an additional covariate in the regression formula \eqref{eq:ag_v}. 

\begin{corollary}\label{cor::comrrALL}
Under Conditions~\ref{cond::1}--\ref{cond::3},
if $\Psi_{vv} = o(1)$, then $\hat \tau_{\HT,\L,\alpha} \mid \mathcal{M}_{\HT}$ is the most peaked around $\tau$ among the set of estimators:
\begin{eqnarray*}
  \{(\hat \tau_{\HT,\L,
  \alpha} \mid \mathcal{M}_{\HT}), \ \hat \tau_{*} , \ (\hat \tau_{*} \mid \mathcal{M}_{*}), \ \hat \tau_{*,\L},\ (\hat \tau_{*,\L} \mid \mathcal{M}_{*}), \ \hat \tau_{*,\P},\ (\hat \tau_{*,\P} \mid \mathcal{M}_{*}): * = \HT, \haj \}.
\end{eqnarray*}
\end{corollary}

Corollary~\ref{cor::comrrALL} establishes the optimality of $\hat \tau_{\HT,\L,\alpha} \mid \mathcal{M}_{\HT}$  among all considered estimators when $\Psi_{vv} = o(1)$, and highlights the utility of including $\alphaw -1$ as an additional covariate in the aggregate regression for ensuring additional efficiency. Intuitively, the unadjusted Hajek estimator $\htau_{\haj}$ implicitly adjusts for the whole-plot sizes, and is hence in general better than the unadjusted Horvitz--Thompson estimator $\htau_{\HT}$; see the comments after Corollary \ref{cor::comrr}.
The $\hat\tau_{\HT, \L,\alpha}$, on the other hand, gives a more efficient way of adjusting for the whole-plot sizes than the Hajek estimator when $\Phi_{vv} = o(1)$. 
  We thus recommend the split-plot rerandomization scheme $\mathcal{M}_{\HT}$ and the associated regression-adjusted estimator $\hat \tau_{\HT,\L,\alpha}$  when the covariates are relatively homogeneous within whole plots or when only whole-plot covariates are used. When the covariates vary greatly within whole plots such that $\Phi_{vv} = o(1)$ does not hold, the projection estimators $\hat \tau_{*,\P} \ (* = \haj,\HT)$ always improve the efficiency under rerandomization, whereas the regression-adjusted estimators $\hat \tau_{\HT,\L,\alpha}$ and $\hat \tau_{*,\L}$ may degrade efficiency compared to the unadjusted counterparts.
This gives an advantage of projection adjustment over regression adjustment. We illustrate this by simulation. 






\section{Numerical Examples}

\subsection{Simulation}
In this section, we conduct simulation to assess the finite-sample performance of the unadjusted and covariate-adjusted estimators under split-plot rerandomization. 
We set $W=600$, $(W_1, W_0) = (0.3W, 0.7W)$,  and generate $(M_{w0}, M_{w1}, M_{w})_{w=1}^W$ as $M_{w0} = \max(2,\zeta_{w0})$, $M_{w1} =\max(2,\zeta_{w1})$, and $M_w = M_{w0} + M_{w1}$, where $\zeta_{w0}$'s are independent Poisson(5) and $\zeta_{w1}$'s are independent Poisson(3). 
For $w=1, \ldots, W$, we draw $v_w = (v_{w1}, v_{w2})^\T$ independently from $\mathcal N((0.6,0.6)^\T, 0.8I_2)$, and use the following two methods to construct subplot covariates {$v_{ws} = (v_{ws,1}, v_{ws,2})^\T$}: (i) $v_{ws}=v_w$ for $s = 1,\ldots,M_w$, which 
corresponds to the case where only whole-plot covariates are used and ensures $\Psi_{vv}= o(1)$; 
(ii) $v_{ws}=v_w + \delta_{ws}$ for $ws \in \mathcal{S}$, where $\delta_{ws}$'s are independent $\mathcal{N} ( 0_2, 0.5I_2 )$, so that the covariates vary within each whole plot. 
{We use $v_{ws}$ for covariate adjustment in the analysis stage, and set $x_{ws} = v_{ws,1}$ for rerandomization in the design stage.} 
The potential outcomes are then generated as
    \[
        \begin{split}
            Y_{ws}(00) &= \theta_w + 0.5 + 2v_{ws,1}^2 + 2v_{ws,2}^2 +\epsilon_{ws},\\
            Y_{ws}(01) &= -0.5\theta_w + 1 + v_{ws,1}^2 + v_{ws,2}^2 +\epsilon_{ws},\\
            Y_{ws}(10) &= 0.5\theta_w + 1 - v_{ws,1}^2 - v_{ws,2}^2 +\epsilon_{ws},\\
            Y_{ws}(11) &= \theta_w + 2 + 2v_{ws,1}^2 + 2v_{ws,1}^2 +\epsilon_{ws}
        \end{split}
    \]
for $ws \in \mathcal{S}$, where $\theta_w$'s are independent $\mathcal{N} ( 2\text{max}(M_w)/M_w, 0.2 )$ and $\epsilon_{ws}$'s are independent Uniform$(-1, 1)$. 
The covariates and potential outcomes are generated once and then kept fixed. 
We perform $2^2$ split-plot randomization and two types of split-plot rerandomization 2,000 times, respectively, and summarize the operating characteristics of $\hts$, $\htau_{*, \L}$,  $\htau_{*, \P}$, and $\htau_{\HT,\L,\alpha}$ for $* = \HT, \haj$.
For rerandomization criteria, we set $d$ to be the 1st percentile 
of $\chi_3^2$, implying an asymptotic acceptance rate of $1\%$. 

Figure~\ref{Simulation1} shows the comparison between estimators under  split-plot randomization and rerandomization when only whole-plot covariates are used. 
The first row illustrates the biases of the covariate-adjusted estimators in finite samples. 
These estimators are asymptotically unbiased, but can have small finite-sample biases \citep{lin2013}.
The second row shows the standard deviations, illustrating the efficiency gain by rerandomization and covariate adjustment. 
Among them, $\hat\tau_{\HT,\L,\alpha}$ under rerandomization is the most efficient, which is coherent with the result of Corollary~\ref{cor::comrrALL}. 
The third row shows the positive empirical biases of standard deviation estimators, implying the conservativeness of distribution estimation. 
The fourth row shows the coverage rates of the constructed $95\%$ confidence intervals, and suggests the validity of all estimators under rerandomization.
The fifth row shows the average confidence interval lengths, which illustrates the efficiency gain by conducting inference with both rerandomization and  covariate adjustment.

Figure \ref{Simulation2} shows the analogous results when covariates vary within each whole plot.
We can see that $\hat\tau_{\HT,\L,\alpha}$ is no longer the most efficient, as $\Psi_{vv} = o(1)$ is not satisfied. 
In this case, the projection estimators $\hat \tau_{*,\P} \ (* = \haj,\HT)$ always improve the efficiency, but the regression-adjusted estimators $\hat \tau_{*,\L}$ may degrade efficiency compared to the unadjusted estimator under  rerandomization. We present an example in the supplementary materials.

\begin{figure}
\includegraphics[width=\linewidth]{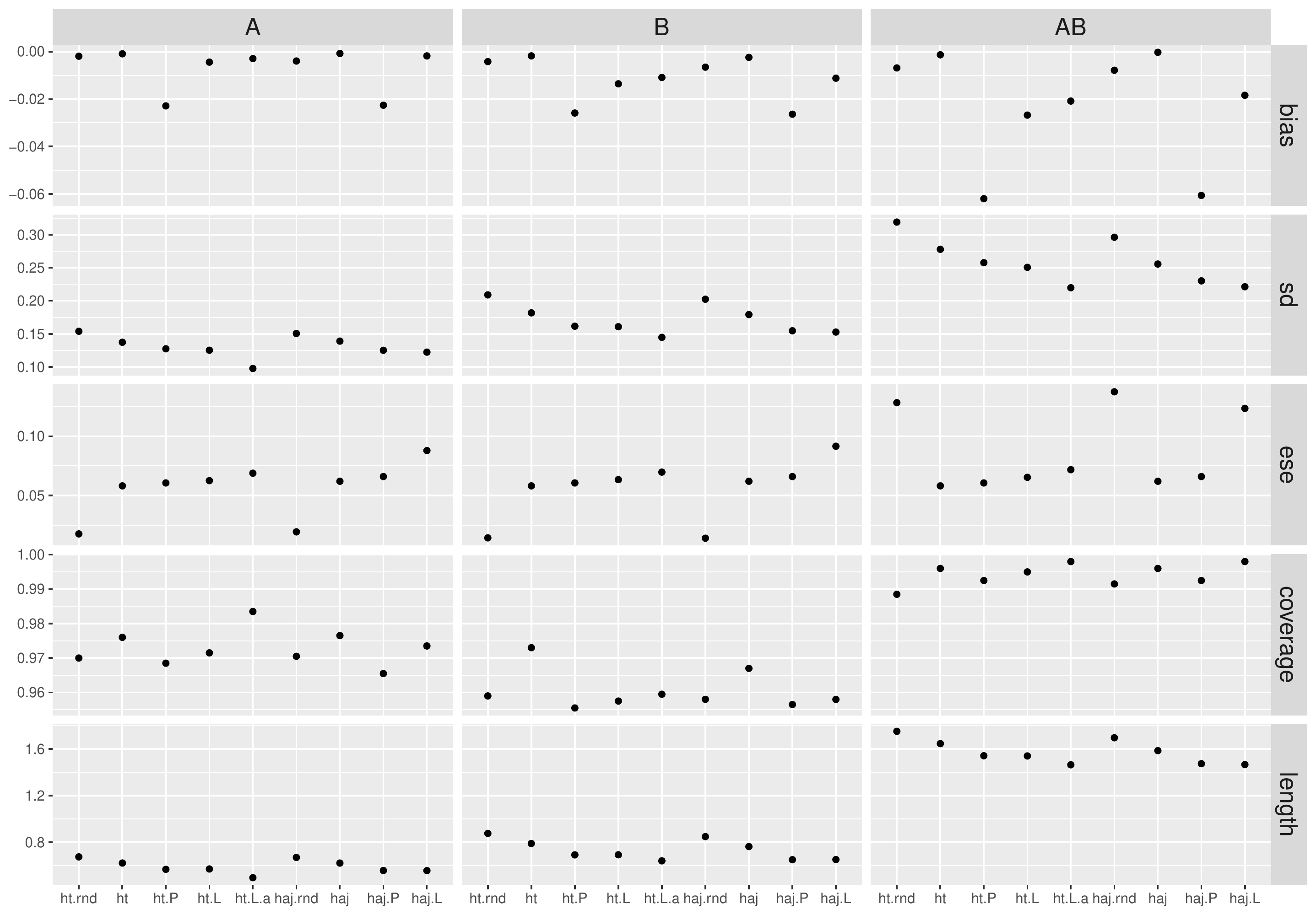}
\caption{
Comparison of the estimators under $2^2$ split-plot randomization and rerandomization with $v_{ws} = v_{w}$. 
The row ``bias" summarizes the average deviations of the point estimators from the true values. The row ``sd" summarizes the standard deviations of the point estimators. 
The row ``ese" summarizes the average errors of the standard deviation estimators. The row ``coverage" summarizes the coverage rates of the 95\% confidence intervals. The row ``length" summarizes the average confidence interval lengths of 95\% confidence intervals. The column ``ht.rnd" stands for $\hat\tau_\HT$ under classic split-plot randomization, ``ht" stands for $\hat\tau_{\HT} \mid \mathcal{M}_{\HT}$, ``ht.P" stands for $\hat\tau_{\HT,\P} \mid \mathcal{M}_{\HT}$, ``ht.L" stands for $\hat\tau_{\HT,\L} \mid \mathcal{M}_{\HT}$, ``ht.L.a" stands for $\hat\tau_{\HT,\L,\alpha} \mid \mathcal{M}_{\HT}$, ``haj.rnd" stands for $\hat\tau_{\haj}$ under classic split-plot randomization, ``haj" stands for $\hat\tau_{\haj} \mid \mathcal{M}_{\haj}$, ``haj.P" stands for $\hat\tau_{\haj,\P} \mid \mathcal{M}_{\haj}$, and ``haj.L" stands for $\hat\tau_{\haj,\L} \mid \mathcal{M}_{\haj}$.
\label{Simulation1}}
\end{figure}

\begin{figure}
\includegraphics[width=\linewidth]{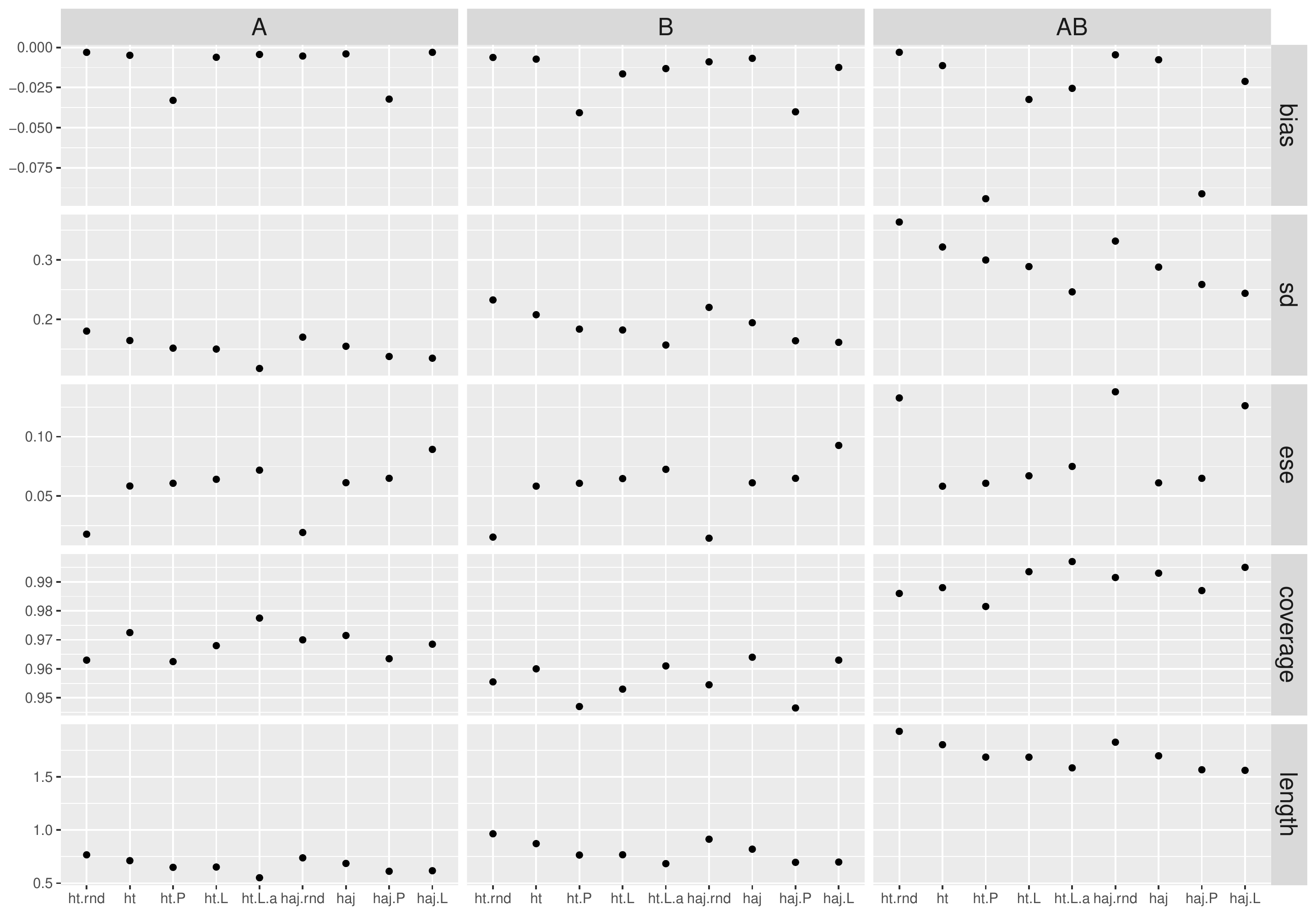}
\caption{Comparison of the estimators under $2^2$ split-plot randomization and rerandomization with varying $v_{ws}$ within each whole plot. The row ``bias" summarizes the average deviations of the point estimators from the true values. The row ``sd" summarizes the standard deviations of the point estimators. The row ``ese" summarizes the average errors of the standard deviation estimators. The row ``coverage" summarizes the coverage rates of the 95\% confidence intervals. The row ``length" summarizes the average confidence interval lengths of 95\% confidence intervals. The column ``ht.rnd" stands for $\hat\tau_\HT$ under classic split-plot randomization, ``ht" stands for $\hat\tau_{\HT} \mid \mathcal{M}_{\HT}$, ``ht.P" stands for $\hat\tau_{\HT,\P} \mid \mathcal{M}_{\HT}$, ``ht.L" stands for $\hat\tau_{\HT,\L} \mid \mathcal{M}_{\HT}$, ``ht.L.a" stands for $\hat\tau_{\HT,\L,\alpha} \mid \mathcal{M}_{\HT}$, ``haj.rnd" stands for $\hat\tau_{\haj}$ under classic split-plot randomization, ``haj" stands for $\hat\tau_{\haj} \mid \mathcal{M}_{\haj}$, ``haj.P" stands for $\hat\tau_{\haj,\P} \mid \mathcal{M}_{\haj}$, and ``haj.L" stands for $\hat\tau_{\haj,\L} \mid \mathcal{M}_{\haj}$. 
\label{Simulation2}}
\end{figure}

\subsection{Real data illustration}
In this section, we analyze a real data set to assess the performance of different estimators under split-plot randomization and rerandomization. \cite{olken2007monitoring} conducted a randomized field experiment on reducing corruption in 608 Indonesian village road projects. We consider two interventions of the study: increasing the probability of external government audits (``audits") and distributing invitations to accountability meetings (``invitations").

The villages are nested in subdistricts, and there was a concern of the spillover effect of audits. 
Therefore, the randomization of audits was clustered by subdistrict. On the other hand, the randomization of invitations was done village by village. This defines a  nonuniform split-plot experiment with the audits and invitations constituting the whole-plot and subplot factors, respectively. 

Before the experiment, \cite{olken2007monitoring} conducted a survey to collect 
ten village characteristics, including village population, village head education, village head salary, and total budget.
To measure the corruption level as the primary outcome of interest, \cite{olken2007monitoring} constructed an independent estimate of the amount each project actually cost to build and then compared it with what the village reported it spent on the project. The percent missing, defined as the difference between the log of the reported amount and the log of the actual amount, is the main measure of corruption level used in the experiment. 

We fill in the missing potential outcomes before the analysis. 
In the data set, there are subdistricts containing only one village, for which we can not calculate covariances within whole plot such as $S_w$ and $S_{w,xY}$. We leave out those subdistricts, and there are 136 subdistricts and 550 villages left. The missing potential outcomes are filled by linear regression based on treatments and ten covariates.

In our analysis, we include village population and village head salary as covariates used for both rerandomization ($x$) 
and 
covariate adjustment ($v$), and focus on the missing percent for materials in road project as the outcome. 
We then perform $2^2$ split-plot randomization and two types of split-plot rerandomization 1,000 times, respectively. For rerandomization criteria, we set $d$ to be the  1st percentile 
of $\chi_6^2$.

Figure~\ref{Olken1} shows the results. Here we use relative standard deviation and average confidence interval length compared to the Horvitz--Thompson estimator under classic split-plot randomization to display the results more clearly. From the second and fifth rows we can see that rerandomization gains estimation and inference efficiency for both the Horvitz--Thompson and Hajek estimators. For example, the standard deviation and average confidence interval length of the Horvitz--Thompson estimator are reduced by rerandomization by approximately $8\%$ for the main effect of audits. Overall, the Hajek estimator performs better than the Horvitz--Thompson estimator. This may be because the 
subdistricts, as the whole plots in our example, have similar average potential outcomes. The covariate-adjusted estimators $\hat \tau_{\HT,\L, \alpha}$, $\hat \tau_{\haj,\L}$, and $\hat \tau_{\haj,\P}$ perform similarly and are the best methods.  


\begin{figure}
\includegraphics[width=\linewidth]{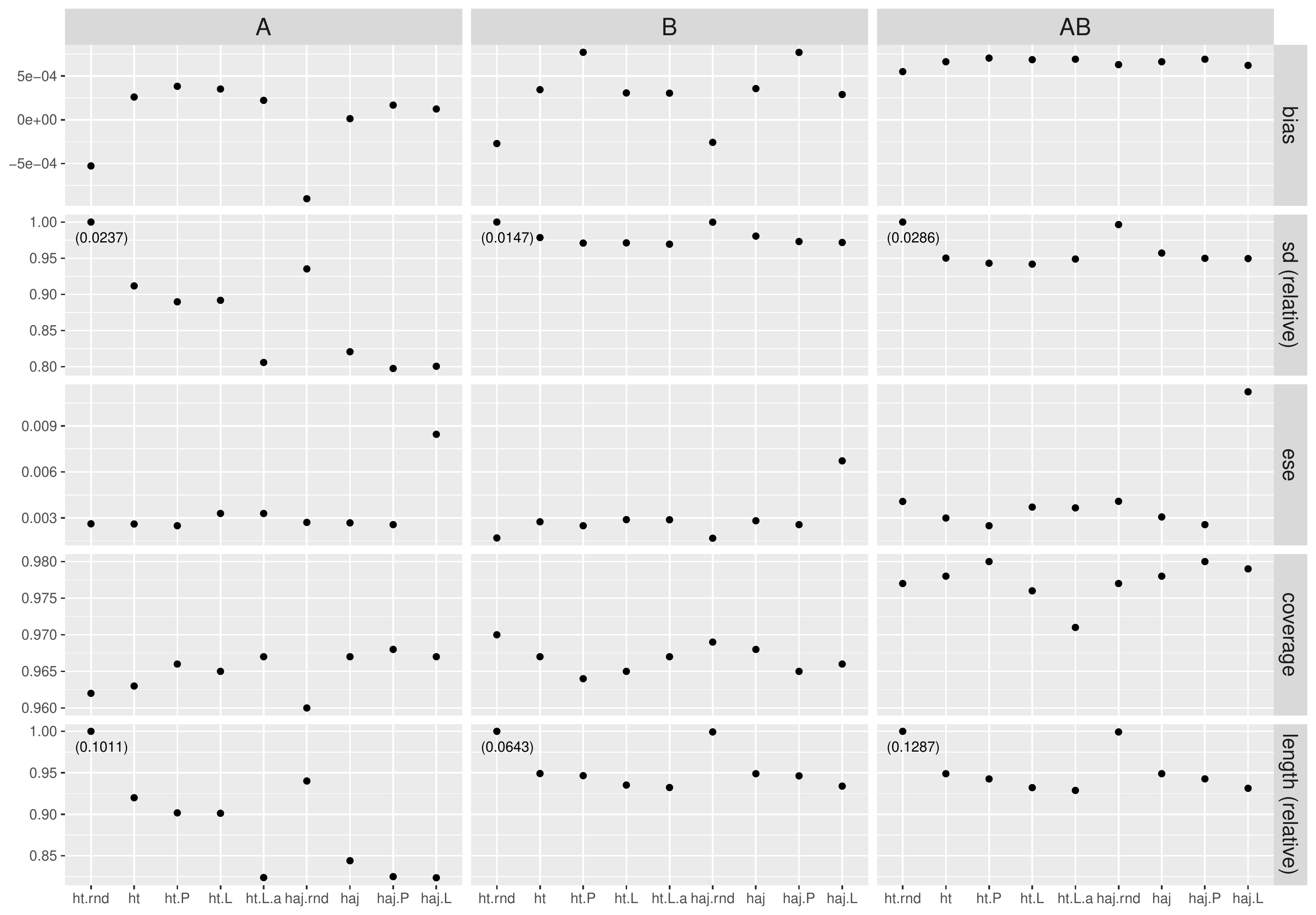}
\caption{ Comparison of different estimators using \cite{olken2007monitoring}'s data set. Factor A and factor B represent audits and invitations, respectively. 
The row ``bias" summarizes the average deviations of the point estimators from the true values. The row ``ese" summarizes the average errors of the standard deviation estimators. The row ``coverage" summarizes the coverage rates of the 95\% confidence intervals. The row ``sd (relative)" and ``length (relative)" summarizes the standard deviations and average confidence interval lengths of 95\% confidence intervals divided by that of Horvitz--Thompson estimator under classic $2^2$ split-plot randomization (``ht.rnd").  The numbers in parentheses are the absolute values for standard deviations and interval lengths. The column ``ht.rnd" stands for $\hat\tau_\HT$ under classic split-plot randomization, ``ht" stands for $\hat\tau_{\HT} \mid \mathcal{M}_{\HT}$, ``ht.P" stands for $\hat\tau_{\HT,\P} \mid \mathcal{M}_{\HT}$, ``ht.L" stands for $\hat\tau_{\HT,\L} \mid \mathcal{M}_{\HT}$, ``ht.L.a" stands for $\hat\tau_{\HT,\L,\alpha} \mid \mathcal{M}_{\HT}$, ``haj.rnd" stands for $\hat\tau_{\haj}$ under classic split-plot randomization, ``haj" stands for $\hat\tau_{\haj} \mid \mathcal{M}_{\haj}$, ``haj.P" stands for $\hat\tau_{\haj,\P} \mid \mathcal{M}_{\haj}$, ``haj.L" stands for $\hat\tau_{\haj,\L} \mid \mathcal{M}_{\haj}$.
\label{Olken1}}
\end{figure}



\section{Discussion}
We investigated the asymptotic properties of rerandomization and covariate adjustment under split-plot designs. 
Based on the asymptotic results, we recommend the use of  rerandomization scheme based on the Horvitz--Thompson estimator if the whole plots have similar total potential outcomes, and  rerandomization scheme based on the Hajek estimator if the whole plots have similar average potential outcomes. 
In the analysis stage, we recommend the fully interacted aggregate regression after adjusting for the whole-plot sizes if only whole-plot covariates are used or more generally, 
$\Psi_{vv} = o(1)$,
and the projection estimator otherwise. The resulting inference is model-free, and remains valid regardless of how well the regression specifications represent the true data generating process of the outcome, treatments, and covariates.

\bigskip
\begin{center}
{\large\bf Supplementary Materials}
\end{center}

\begin{description}
\item The supplementary materials provide additional simulation results and proofs.
\end{description}

\bibliographystyle{agsm}

\bibliography{paper-ref_0921}



\newpage

\def\spacingset#1{\renewcommand{\baselinestretch}%
{#1}\small\normalsize} \spacingset{1}

\renewcommand\thesection{S\arabic{section}}
\renewcommand\thesubsection{S\arabic{section}.\arabic{subsection}}
\setcounter{section}{0}
\renewcommand\thelemma{S\arabic{lemma}}
\setcounter{lemma}{0}

\if0\blind
{
  \title{\bf Supplementary material for ``Rerandomization and covariate adjustment in spilt-plot designs"}
  \author{Wenqi Shi\\
    Department of Industrial Engineering, Tsinghua University, \\
    Beijing, 100084, China\vspace{.2cm}\\
    Anqi Zhao\thanks{Zhao was supported by the Start-Up grant R-155-000-216-133 from the National University of Singapore.}\\
    Department of Statistics and Data Science, National University of Singapore, \\
    117546, Singapore\vspace{.2cm}\\
    Hanzhong Liu\thanks{Liu was supported by the National Natural Science Foundation of China (12071242).}\hspace{.2cm}\\
    Center for Statistical Science, Department of Industrial Engineering, \\
    Tsinghua University, Beijing, 100084, China}
  \date{}
  \maketitle
} \fi

\if1\blind
{
  \bigskip
  \bigskip
   \bigskip
   \begin{center}
     {\LARGE\bf Supplementary Material for ``Rerandomization and covariate adjustment in spilt-plot designs"}
 \end{center}
   \medskip
} \fi



\spacingset{1.9} 

Section~\ref{sec:A} gives additional simulation results. Section~\ref{sec:B} gives the proofs.


\section{Additional simulation results}\label{sec:A}

In this section, we raise an extreme case to show that estimators adjusted by Lin's method can be less efficient than the unadjusted estimator under corresponding rerandomization schemes. We set $W=1200$, $(W_1, W_0) = (0.9W, 0.1W)$, and generate $(M_{w0}, M_{w1}, M_{w})_{w=1}^W$ as $M_{w0} =$ max $(2,\zeta_{w0})$, $M_{w1} =$ max $(2,\zeta_{w1})$, and $M_w = M_{w0} + M_{w1}$, where $\zeta_{w0}$'s are independent Poisson(3) and $\zeta_{w1}$'s are independent Poisson(8). For $w=1, \ldots, W$, we still draw independently whole-plot average covariates $v_w$ from $N((0.6,0.6)^\T, 0.8I_2)$, but covariates are more varying within whole-plots by setting $v_{ws}=v_w + \delta_{ws}$, where $\delta_{ws}$'s are independent $\mathcal{N} ( 0_2, 2I_2 )$. We set covariates $x_{ws} = v_{ws}$, which means rerandomization and covariate adjustments use the same information. The potential outcomes are then generated as
\begin{eqnarray*}
Y_{ws}(00) &=& \theta_w + 0.5 + 2\bar v_{w1}^2 + 2\bar v_{2}^2 +\epsilon_{ws}, \\
Y_{ws}(01) &=& -0.5\theta_w + 1 + \bar v_{w1}^2 + \bar v_{2}^2 +\epsilon_{ws}, \\
Y_{ws}(10) &=& 0.5\theta_w + 1 - \bar v_{w1}^2 - \bar v_{2}^2 +\epsilon_{ws},\\
Y_{ws}(11) &=& \theta_w + 2 + 2\bar v_{w1}^2 + 2\bar v_{2}^2 +\epsilon_{ws},
\end{eqnarray*}
for $ws \in \mathcal{S}$, where $\theta_w$'s are indepedent $\mathcal{N} ( 2\text{max}(M_w)/M_w, 0.2 )$ and $\epsilon_{ws}$'s are independent Unif(-1, 1). Here, $\bar v_{w1}$ and $\bar v_{2}$ denote the first element of whole-plot averaged covariates and the second element of covariates averaged over the whole population. For rerandomization criteria, we set $d$ to be the 0.01 quantile of $\chi_6^2$, so that the asymptotic acceptance rate is 0.01. We use only estimators based on Horvitz--Tompson method. 

The result is summarized in Fig~\ref{Simulation3}. Because the standard deviations of the estimators for the main effects and interaction have very different scaling, we use relative standard deviations and average confidence interval lengths compared to the Horvitz--Tompson estimator under classical split-plot randomization to display the results more clearly. We can see that Lin's regression-adjusted estimators cannot guarantee efficiency gain compared to the unadjusted estimator under rerandomization, while the adjustment methods based on the projection or conditional inference perspective can still guarantee efficiency improvement.


\begin{figure}
\includegraphics[width=\linewidth]{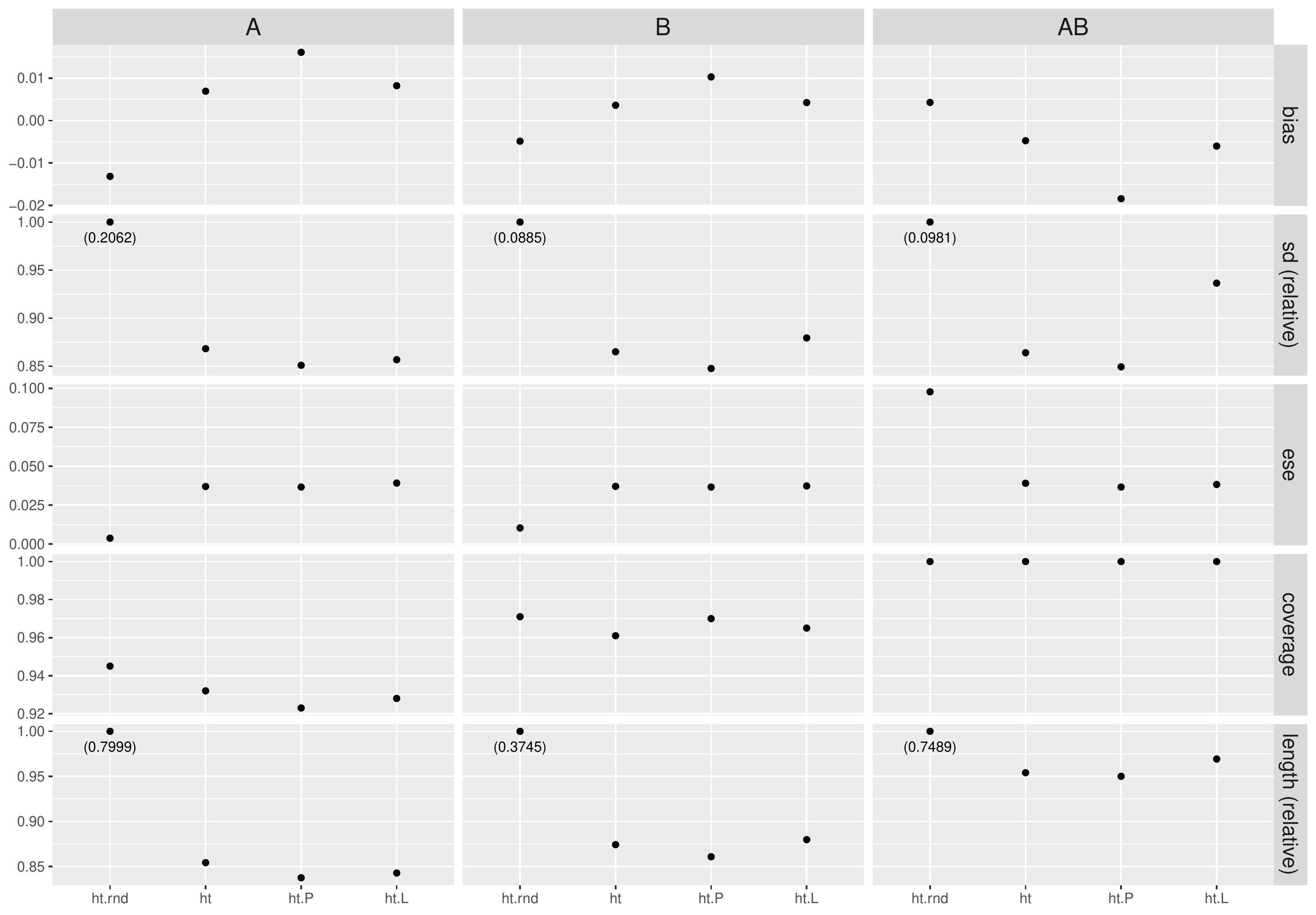}
\caption{The possible efficiency decrease of estimators adjusted by Lin's method. The row ``bias" summarizes the average deviations of the point estimators from the true values. The row ``ese" summarizes the average errors of the standard deviation estimators. The row ``coverage" summarizes the coverage rates of 95\% confidence intervals. The row ``sd (relative)" and ``length (relative)" summarizes the standard deviations and average confidence interval lengths of 95\% confidence intervals divided by that of Horvitz--Tompson estimator under classical split-plot randomization (``ht.rnd"). The column ``ht.rnd" stands for $\hat\tau_\HT$ under classic split-plot randomization, ``ht" stands for $\hat\tau_{\HT} \mid \mathcal{M}_{\HT}$, ``ht.P" stands for $\hat\tau_{\HT,\P} \mid \mathcal{M}_{\HT}$, ``ht.L" stands for $\hat\tau_{\HT,\L} \mid \mathcal{M}_{\HT}$. The numbers in parentheses are the absolute values for standard deviations and average confidence interval lengths.
\label{Simulation3}}
\end{figure}

\section{Proofs}\label{sec:B}

\subsection{Proof of Theorem~\ref{thm::clt}}

Before the proof, we give the explicit formulas of $S_{*,xx}$ and $S_{*,xY(z)}$. For $* = \HT, \haj$, $S_{*,xx}$ and $S_{w,xx}$ are defined as
\begin{eqnarray*}
S_{\HT,xx} &=&  (W-1)^{-1}\sumw \left( \alphaw \bar x_{w} - \bar{x}  \right) \left( \alpha_w \bar x_{w} - \bar{x} \right)^\T, \\
S_{\haj,xx} &=& (W-1)^{-1}\sumw \alphaw^2 \left( \bar x_{w} - \bar{x} \right) \left(  \bar x_{w} - \bar{x} \right)^\T, \\
S_{w,xx} &=& (M_w-1)^{-1} \sums \alphaw^2 \left( x_{ws}  - \bar x_{w} \right)\left(  x_{ws} - \bar x_{w}  \right)^\T, 
\end{eqnarray*}
and $S_{*,xY(z)}$ and $S_{w,xY(z)}$ are defined as
\begin{eqnarray*}
S_{\HT,xY(z)} &=&  (W-1)^{-1}\sumw  \left( \alpha_w \bar x_{w} - \bar{x} \right) \left\{ \alphaw \bar Y_{w}(z) - \bar{Y}(z)  \right\}, \\
S_{\haj,xY(z)} &=& (W-1)^{-1}\sumw \alphaw^2 \left( \bar x_{w} - \bar{x} \right) \left\{  \bar Y_{w}(z) - \bar{Y}(z) \right\},\\
S_{w,xY(z)} &=& (M_w-1)^{-1} \sums \alphaw^2 \left(  x_{ws} - \bar x_{w}  \right) \left\{ Y_{ws}(z)  - \bar Y_{w}(z) \right\}.
\end{eqnarray*}


Our proof relies on the finite-population central limit theory for $\hat Y_*$ with scalar potential outcomes $Y_{ws}(z)$ under the $2^2$ split-plot randomization \citep[][Theorem 1]{zhao2021reconciling}; see Lemma~\ref{lemma::clt} below.  


\begin{lemma} 
\label{lemma::clt}
Under Condition~\ref{cond::1}, for $* = \HT, \haj$,
$$
    \sqrt{W}(\hat Y_{*}-\bar{Y}) \rs N(0,V_{*,YY}),
$$
where $V_{*,YY} = H \otimes S_* + \Psi$.
\end{lemma}


We extend Lemma~\ref{lemma::clt} to the joint asymptotic distribution of $\hat Y_*$ and $\hat x_*$ under Conditions~\ref{cond::1}--\ref{cond::2} by showing that Lemma~\ref{lemma::clt} applies to any linear combination of  $\hat Y_*$ and $\hat x_*$.

In the proof below, let $||\cdot||_1$ and $||\cdot||_2$  denote the $\ell_1$ and $\ell_2$ norms, respectively. Write $U_1 \sim U_2$ if random variables $U_1$ and $U_2$ have the same distribution. Without loss of generality, we assume that the covariates are centered such that $\bar{x} = N^{-1} \sumws x_{ws} = 0$ and $\bar{v} = N^{-1} \sumws v_{ws} = 0$.


\begin{proof}[Proof of Theorem~\ref{thm::clt}]
Denote
$$
V_{*} = \left(\begin{array}{cc}
        V_{*, YY} & V_{*, Yx} \\  
        V_{*, xY} & V_{*, xx}  \\
   \end{array} \right)
$$
with $V_{*,xY} = V_{*,Yx}^\T = (H\otimes 1_L)\circ(1_4\otimes S_{*,xY}) + \Psi_{xY}$ and $V_{*,xx} = H\otimes S_{*,xx}+\Psi_{xx}$. It suffices to show that
\begin{eqnarray}
   \sqrt{W} \left( \begin{array}{cc}
        \hat Y_* - \bar Y  \\
        \hat x_* 
   \end{array} \right) \rs 
   \mathcal{N} \left( 0, V_{*} \right). \nonumber
\end{eqnarray}
The joint asymptotic normality of $\hat Y_* - \bar Y$ and  $\hat x_*$ can be obtained by showing that their linear combinations are asymptotically normal. That is, it suffices to show that, for any 
fixed $u = (u_y^\T , u_x^\T)^\T \in \mathbb R^{4 (1 + L)}$ with 
$$
u_y = (u_{y}(00), u_{y}(01), u_{y}(10), u_{y}(11))^\T, \quad u_x = (u_{x}(00)^\T, u_{x}(01)^\T, u_{x}(10)^\T, u_{x}(11)^\T)^\T,
$$
and $||u||_2 = 1$, the linear combination $u_y^\T (\hat Y_* - \bar Y )  + u_x^\T \hat x_* $ is asymptotically normal with mean zero and covariance $u^\T V_{*} u$. Note that
\begin{eqnarray}
  u_y^\T (\hat Y_* - \bar Y )  + u_x^\T \hat x_* & = & \sum_{z \in \mathcal{T}} \big[ u_y(z) \{ \hat Y_{*}(z) - \bar{Y}(z) \} + u_x(z)^\T \hat x_{*}(z) \big]. \nonumber
\end{eqnarray}
Define the transformed outcome $R_{ws}(z) = u_y(z) Y_{ws}(z) + u_x(z)^\T x_{ws} $. Let $\Psi(z,z^\prime)$ be the element of $\Psi$ corresponding to $(z,z^\prime)$. Define $\hat R_{*}$, $\bar R(z)$, $\bar R_w(z)$, $\overline{R_{w.}^4 (z)}$, $S_{*,R}(z,z^\prime)$, and $\Psi_R(z, z^\prime)$ similarly to $\hat Y_{*}$, $\bar Y(z)$, $\bar Y_w(z)$, $\overline{Y_{w.}^4 (z)}$, $S_{*}(z,z^\prime)$, and $\Psi(z, z^\prime)$ with $Y_{ws}(z)$ replaced by $R_{ws}(z)$. Then $u_y^\T (\hat Y_* - \bar Y )  + u_x^\T \hat x_*$ is the linear combination (summation) of the components of $\hat R_{*}$. By Lemma~\ref{lemma::clt}, it suffices for the asymptotic normality of $u_y^\T (\hat Y_* - \bar Y )  + u_x^\T \hat x_*  $ to show that $R_{ws}(z)$'s satisfy Condition~\ref{cond::1}. Since Condition~\ref{cond::1}(i)-(ii) are satisfied naturally, we only need to show that  Condition~\ref{cond::1}(iii)--(v) hold for $R_{ws}(z)$'s.

For (iii), since $\bar{x}=0$, simple calculation gives
\begin{eqnarray}
  \bar{R}(z) &=& u_y(z) \bar{Y}(z), \nonumber \\
  S_{*,R} (z, z^\prime) &= & u_y(z) S_* (z, z^\prime) u_y(z^\prime) + u_x(z)^\T S_{*,xx} u_x(z^\prime) + u_y(z) S^\T_{*, xY(z)} u_x(z^\prime) \nonumber \\
  && +  u_x(z)^\T S_{*,x Y(z^\prime)}  u_y(z^\prime), \nonumber \\
  {\Psi}_R(z, z^\prime) &=& u_y(z) \Psi (z, z^\prime) u_y(z^\prime) + u_x(z)^\T  \Psi_{xx} u_x(z^\prime) + u_y(z) \Psi_{ Y(z)x} u_x(z^\prime)  \nonumber \\
  && + u_x(z)^\T  \Psi_{x Y(z^\prime)} u_y(z^\prime). \nonumber
\end{eqnarray}
Here, $\Psi_{xY(z)} = \Psi_{Y(z)x}^\T$ is the column of $\Psi_{x Y}$ corresponding to treatment $z$
(the $1-4$ columns correspond to $z=(00),(01),$ $(10),(11)$).
Thus, $S_{*,R}$, $\bar{R}$, and ${\Psi}_R$ have finite limits (note that the asymptotic normality still holds if the limit of $\Sigma_{*, \tau \tau}$ is not invertible).

For (iv), we have
\begin{eqnarray*}
& & W^{-1} \max_{w=1,...,W} \; | \alpha_w \bar{R }_w(z)-\bar{R}(z)|^2\\
&=& W^{-1}  \max_{w=1,...,W}\; \big[ u_y(z)^2 \{ \alpha_w \bar{{Y} }_w(z)-\bar{{Y}}(z)\}^2 +  u_x(z)^\T ( \alpha_w \bar{{x} }_w-\bar{{x}}) ( \alpha_w \bar{{x} }_w-\bar{{x}})^\T u_x(z) \\ & &+ 2 u_y(z)  \{ \alpha_w \bar{{Y} }_w(z)-\bar{{Y}}(z)\} ( \alpha_w \bar{{x} }_w-\bar{{x}})^\T u_x(z)  \big] \\
&\leq & 2 W^{-1} \big[ u_y(z)^2  \max_{w=1,...,W}  \{\alpha_w \bar{Y}_{w}(z)-\bar{Y}(z)\}^2 + ||u_x(z)||_1^2  \max_{w=1,...,W} || \alpha_w \bar{x}_{w} - \bar{x}||_\infty^2  \big]  \\
&=& o(1).
\end{eqnarray*}

For (v), we have
\begin{eqnarray*}
W^{-1} \sum_{w=1}^W \alpha_w^2 \overline{R_{w.}^4 (z)} &\leq &
W^{-1}\sum_{w=1}^W8\alpha_w^2\Big\{ u^4_y(z) \overline{ Y_{w.}^4(z) }+ L ||u_x(z)||_\infty^4 \overline{|| x_{w.} ||_\infty^4}  \Big\}  = O(1), \\
W^{-2} \sum_{w=1}^W \alpha_w^4 \overline{R_{w.}^4 (z)} & \leq & W^{-2}\sum_{w=1}^W8\alpha_w^4\{ u^4_y(z) \overline{ Y_{w.}^4(z) } + L ||u_x(z)||_\infty^4 \overline{|| x_{w.} ||_\infty^4} \} = o(1).
\end{eqnarray*}

\end{proof}

\subsection{Proof of Theorem~\ref{thm::rerand}}


\begin{proof}[Proof of Theorem~\ref{thm::rerand}]
Let $\epsilon \sim \mathcal{N}(0, I_{3})$ be a 3-dimensional standard normal random vector, and $D=(D_1,...D_{3L})^\T \sim \mathcal{N}(0,I_{3L})$ be a $3L$-dimensional standard normal random vector, independent of $\epsilon$. Denote $\sqrt{W} \tilde\tau_* = (\Sigma_{*,\tau\tau}^{\bot})^{1/2}\epsilon + \Sigma_{*,\tau x}\Sigma_{*,xx}^{-1/2} D$ and $\sqrt{W} \tilde \tau_{*,x} = \Sigma_{*,xx}^{1/2}D$. Then
\begin{eqnarray*}
 \sqrt{W} \left( \begin{array}{cc}
        \tilde\tau_*  \\
        \tilde \tau_{*,x}
  \end{array} \right) \sim 
  \mathcal{N} \left( 0, \Sigma_* \right).
\end{eqnarray*}
By Theorem~\ref{thm::clt} and \citet[][Proposition A1]{Li2018},
\begin{eqnarray}
  \sqrt{W} \left( \begin{array}{cc}
        \hat \tau_* - \tau \\
        \hat \tau_{*,x}
  \end{array} \right)  \mid \hat \tau_{*,x}^\T \cov( \hat \tau_{*,x} )^{-1} \hat \tau_{*,x} \leq d & \rs &
  \sqrt{W} \left( \begin{array}{cc}
        \tilde\tau_*  \\
        \tilde \tau_{*,x}
  \end{array} \right) \mid \tilde \tau_{*,x}^\T \cov( \tilde \tau_{*,x} )^{-1} \tilde \tau_{*,x} \leq d. \nonumber
\end{eqnarray}
Note that the above conclusion holds if we replace $\cov(\cdot)$ by $\cova(\cdot)$. Then, for $* = \HT, \haj$, 
\begin{eqnarray*}
  \sqrt{W}(\hat\tau_{*}- \tau) \mid \mathcal{M}_{*} & \rs &
  \sqrt{W} \tilde\tau_* \mid W \tilde \tau_{*,x}^\T \Sigma_{*,xx}^{-1} \tilde \tau_{*,x} \leq d \\
  & \sim & (\Sigma_{*,\tau\tau}^{\bot})^{1/2}\epsilon + \Sigma_{*,\tau x}\Sigma_{*,xx}^{-1/2} D \mid D^\T D \leq d\\
  & \sim & (\Sigma_{*,\tau\tau}^{\bot})^{1/2}\epsilon
            +\Sigma_{*,\tau x}\Sigma_{*,xx}^{-1/2}\zeta_{3L, d}.
\end{eqnarray*}
\end{proof}


\subsection{Proof of Corollary~\ref{cor::varcom}}


First, we introduce without proof a few lemmas obtained by \citet{Morgan2012} and \citet{Li2020factorial}.

\begin{lemma}
\label{lemma::Lcov}
$\cov(\zeta_{3L, d}) = r_{3L, d} I_{3L}$, where $r_{3L, d} =  \pr( \chi^2_{3L + 2} \leq d )/ \pr( \chi^2_{3L} \leq d ) $.
\end{lemma}

We write $\phi \succ \varphi$ if for every  symmetric convex set $\mathcal{K} \in \mathbb{R}^m$, $\pr (\phi \in \mathcal{K}) \geq \pr (\varphi \in \mathcal{K}) $. Lemmas~\ref{lemma::peak1} and ~\ref{lemma::peak2} below provide useful results for peakness comparison.

\begin{lemma} \label{lemma::peak1}
If two $m$ dimensional symmetric random vectors $\phi_1$ and $\phi_2$ satisfy $\phi_1 \succ \phi_2$, then for any non-random matrix $C \in \mathbb{R}^{p \times m}$, $C\phi_1 \succ C\phi_2$.
\end{lemma}

\begin{lemma} \label{lemma::peak2}
Let $\phi_1$, $\phi_2$ and $\varphi$ be three symmetric random vectors; $\phi_1$ and $\varphi$, $\phi_2$ and $\varphi$ are independent. If $\phi_1 \succ \phi_2$ and $\varphi$ is central symmetric unimodal, then $\varphi + \phi_1 \succ \varphi + \phi_2$.
\end{lemma}

\begin{lemma}\label{lemma::ccu}
If $\phi \in \mathbb{R}^m$ is central convex unimodal, then for any non-random matrix $C \in \mathbb{R}^{p \times m}$, $C \phi \in \mathbb{R}^{p}$ is also central convex unimodal.
\end{lemma}

\begin{lemma} \label{lemma::peak3}
For $\zeta_{3L, d}\sim D | D^\T D \leq d$ with $D=(D_1,...D_{3L})^\T \sim \mathcal{N}(0,I_{3L})$, $\zeta_{3L, d} \succ D$.
\end{lemma}

\begin{proof}[Proof of Corollary~\ref{cor::varcom}]

For $* = \HT, \haj$, by Theorem~\ref{thm::rerand} and Lemma~\ref{lemma::Lcov}, we have 
\begin{eqnarray*}
   W \cova (\hat\tau_{*} \mid \mathcal{M}_{*}) &=&
   \Sigma_{*,\tau\tau}^{\bot} + \Sigma_{*,\tau x} \Sigma_{*,xx}^{-1/2} \cova(\zeta_{3L, d}) \Sigma_{*,xx}^{-1/2} \Sigma_{*,x \tau}\\
   &=&\Sigma_{*,\tau\tau}^{\bot} + r_{3L, d} \Sigma_{*,\tau\tau}^{||}\\
   &=& \Sigma_{*,\tau\tau} - (1-r_{3L, d})\Sigma_{*,\tau\tau}^{||}.
\end{eqnarray*}

Since $0 \leq r_{3L, d} \leq 1$,  $\Sigma_{*,\tau\tau}^{||}$ is positive semi-definite, and $W\cova (\hat\tau_{*}) = \Sigma_{*,\tau\tau}$,
$$
W \left[\cova( \hat \tau_{*}) - \cova( \hat \tau_{*} \mid \mathcal{M}_{*}) \right] = ( 1 - r_{3L, d}) \Sigma_{*, \tau\tau}^{||} \geq 0.
$$

By Lemmas~\ref{lemma::peak1} and \ref{lemma::peak3}, $\Sigma_{*,\tau x}\Sigma_{*,xx}^{-1/2}\zeta_{3L, d} \succ \Sigma_{*,\tau x}\Sigma_{*,xx}^{-1/2}D$. We can derive from Lemma~\ref{lemma::ccu} that $(\Sigma_{*,\tau\tau}^{\bot})^{1/2}\epsilon$ is central convex unimodal, which, coupled with Lemma~\ref{lemma::peak2}, ensures that $(\Sigma_{*,\tau\tau}^{\bot})^{1/2} \epsilon  + \Sigma_{*,\tau x} \Sigma_{*,xx}^{-1/2} \zeta_{3L, d} \succ (\Sigma_{*,\tau\tau}^{\bot})^{1/2} \epsilon  + \Sigma_{*,\tau x} \Sigma_{*,xx}^{-1/2} D$. Recall that $$\sqrt{W} (\hat \tau_{*} - \tau) \rs  \mathcal{N}(0, \Sigma_{*,\tau\tau}) \sim  (\Sigma_{*,\tau\tau}^{\bot})^{1/2} \epsilon  + \Sigma_{*,\tau x} \Sigma_{*,xx}^{-1/2} D.$$
Hence, $\pra\{ \sqrt{W} (\hat \tau_{*} - \tau ) \in \mathcal{K} \mid \mathcal{M}_{*} \} \geq \pra \{ \sqrt{W} (\hat \tau_{*} - \tau ) \in \mathcal{K} \}$ for every symmetric convex set $\mathcal{K} \subset \mathbb{R}^3$. That is, rerandomization by $\mathcal{M}_{*}$ improves the asymptotic efficiency of $\hts$.

\end{proof}

\subsection{Proof of Theorem~\ref{thm::varest}}


\begin{lemma}\label{lem::sample-covariance}
Under Conditions~\ref{cond::1}--\ref{cond::2}, for $* = \HT,\haj$, $\hat \Sigma_{*,x\tau} - \Sigma_{*,x\tau} = o_\mathbb P(1)$.
\end{lemma}
\begin{proof}[Proof of Lemma~\ref{lem::sample-covariance}]
It suffices to show that $E(\hat \Sigma_{*,x\tau}) = \Sigma_{*,x\tau}$ and $\cov(\hat \Sigma_{*,x\tau}) =o(1)$ as $W$ goes to infinity. We first prove the unbiasedness of the estimators. For all $z \in \mathcal{T}$, the Horvitz–-Thompson estimators of $\bar{Y}(z)$, $\bar{Y}_w(z)$ and $Y_{ws}(z)$ are unbiased. That is,
\begin{eqnarray*}
E\{ \hat Y_{\HT}(z) \} & = & \bar{Y}(z), \\
E \{ \hat Y_{\HT,w}(z) \} &=& 
M_{w}^{-1}E \Big\{ \sum_{ws \in \mathcal{S}(z)} p_{ws}(z)^{-1}Y_{ws}(z) \Big\}
= M_{w}^{-1} \sum_{s=1}^{M_w} Y_{ws}(z) = \bar{Y}_w(z),\\
E \{ \hat Y_{\HT,ws}(z) \} &=&  E \{ \mathcal{I}(Z_{ws}=z) p_{ws}(z)^{-1} Y_{ws}(z) \} = Y_{ws}(z).
\end{eqnarray*}
Thus,
$$
E\left[\hat S_{\HT,xY(z)}\right] = E\Big[ (W - 1)^{-1}\sumw \left( \alpha_w \bar x_{w} - \bar{x} \right) \left\{ \alphaw \hat Y_{\HT,w}(z) - \hat{Y}_{\HT}(z)   \right\} \Big]\\
= S_{\HT,xY(z)}.
$$
Similarly, $E(\hat S_{w,xY(z)}) = S_{w,xY(z)}$ and $E(\hat S_{\haj,xY(z)}) = S_{\haj,xY(z)}$. Therefore, $E(\hat \Sigma_{*,x\tau}) = \Sigma_{*,x\tau}$.

Note that $\bar x = 0$ and $\hat Y_{\HT}(z) = W^{-1} \sumw \alphaw \hat Y_{\HT,w}(z)$. Denote $R_w = \alphaw^2 \bar{x}_w \hat Y_{\HT,w}(z)$ to write $\hat S_{\HT,xY(z)} = (W-1)^{-1} \sumw R_w$ for $z=(ab)$. To bound $$\cov( \hat S_{\HT,xY(z)}) = (W-1)^{-2} \Big\{\sumw \cov(R_w) + \sum_{w \neq k} \cov( R_w, R_k ) \Big\},$$ for $w \neq k$, we have
\begin{eqnarray*}
    E(R_w\mid A_w=a) & = & p_a^{-1} \alphaw^2 \bar{x}_w \bar{Y}_w(z),\\
    E\{ \cov (R_w \mid A_w) \} & = & p_a \cov(R_w \mid A_w = a) \\
    &=& p_a E(R_w R_w^\T \mid A_w=a) - p_a^{-1} \alphaw^4 \bar{x}_w \bar{x}_w^\T \bar{Y}^2_w(z),\\
    \cov\{ E(R_w \mid A_w)  \} &=& E[E(R_w \mid A_w) E(R_w \mid A_w)^\T] - E(R_w) E(R_w)^\T \\
    &=& p_a^{-1} \alphaw^4 \bar{x}_w \bar{x}_w^\T \bar{Y}^2_w(z) - \alphaw^4 \bar{x}_w \bar{x}_w^\T \bar{Y}^2_w(z),\\
    E\{ E(R_w \mid A_w) E(R_k \mid A_k)^\T \} &=& \pr (A_w = A_k = a) E(R_w \mid A_w=a)E(R_k \mid A_k=a)^\T\\
    &=& p_a \frac{W_a -1}{W-1}p_a^{-2} \alphaw^4 \bar{x}_w \bar{x}_k^\T \bar{Y}_w(z)\bar{Y}_k(z),\\
    E(R_w)E(R_k)^\T &=& \alphaw^4 \bar{x}_w \bar{x}_k^\T \bar{Y}_w(z)\bar{Y}_k(z),\\
    E\{\cov(R_w, R_k \mid A_w, A_k)\} &=& \pr(A_w = A_k = a) \cov(R_w, R_k \mid A_w = A_k = a) = 0.
\end{eqnarray*}
Hence,
\begin{eqnarray*}
    \cov(R_w) &=&  E\{ \cov(R_w \mid A_w) \} + \cov\{ E(R_w \mid A_w) \}\\
    &=& p_a E(R_w R_w^\T \mid A_w=a) - \alphaw^4 \bar{x}_w \bar{x}_w^\T \bar{Y}^2_w(z),\\
    \cov(R_w, R_k) &=& \cov\{E(R_w\mid A_w), E(R_k\mid A_k)\} + E\{\cov(R_w, R_k \mid A_w, A_k)\}\\
    &=& E\{ E(R_w \mid A_w) E(R_k \mid A_k) \} - 
    E(R_w)E(R_k)\\
    &=& -p_1p_0(W-1)^{-1}p_a^{-2} \alphaw^4 \bar{x}_w \bar{x}_k^\T \bar{Y}_w(z) \bar{Y}_k(z).
\end{eqnarray*}
This ensures that
\begin{eqnarray*}
    && (W-1)^{2}\cov( \hat S_{\HT,xY(z)}) 
    =  \sumw \cov(R_w) + \sum_{w \neq k} \cov( R_w, R_k )\\
    &=&  \sumw \left[ p_a E(R_w R_w^\T \mid A_w=a) - \alphaw^4 \bar{x}_w \bar{x}_w^\T \bar{Y}^2_w(z) \right] - p_1p_0(W-1)^{-1}p_a^{-2} \sum_{w \neq k} \alphaw^4 \bar{x}_w \bar{x}_k^\T \bar{Y}_w(z) \bar{Y}_k(z)\\
    &=& p_a \sumw E(R_w R_w^\T \mid A_w=a) -
    p_1p_0(W-1)^{-1}p_a^{-2} \sum_{w, k} \alphaw^4 \bar{x}_w \bar{x}_k^\T \bar{Y}_w(z) \bar{Y}_k(z)\\
    && - p_a^{-2} \{ p_a^2 - p_1p_0(W-1)^{-1} \} \sumw \alphaw^4 \bar{x}_w \bar{x}_w^\T \bar{Y}^2_w(z)\\
    &\leq&  p_a \sumw E(R_w R_w^\T \mid A_w=a) - p_a^{-2} \{ p_a^2 - p_1p_0(W-1)^{-1} \} \sumw \alphaw^4 \bar{x}_w \bar{x}_w^\T \bar{Y}^2_w(z).
\end{eqnarray*}
Therefore, $\cov( \hat S_{\HT,xY(z)})$ is bounded by $(W-1)^{-2} p_a \sumw E(R_w R_w^\T \mid A_w=a) = o(1)$ as $W$ goes to infinity. Given $\cov( \hat S_{\HT,xY(z)}) = o(1)$ and $E( \hat S_{\HT,xY(z)}) =  S_{\HT,xY(z)}$, Markov’s inequality ensures that $\hat S_{\HT,xY(z)} - S_{\HT,xY(z)} = o_\mathbb P(1)$. Similarly, $\hat S_{\haj,xY(z)} - S_{\haj,xY(z)} = o_\mathbb P(1)$. 

Let $H_w(z,z^\prime)$ be the element of $H_w$ corresponding to $(z,z^\prime)$. Denote $$\Psi_{xY}(z,z^\prime) = W^{-1}\sumw M_w^{-1} H_w(z,z^\prime)S_{w,xY(z^\prime)} \in \mathbb{R}^{L}$$ to write
$$
    \Psi_{xY} =  \left( \begin{array}{cccc}
        \Psi_{xY}(00,00) & \Psi_{xY}(00,01) & \Psi_{xY}(00,10) & \Psi_{xY}(00,11)  \\
        \Psi_{xY}(01,00) & \Psi_{xY}(01,01) & \Psi_{xY}(01,10) & \Psi_{xY}(01,11)  \\
        \Psi_{xY}(10,00) & \Psi_{xY}(10,01) & \Psi_{xY}(10,10) & \Psi_{xY}(10,11)  \\
        \Psi_{xY}(11,00) & \Psi_{xY}(11,01) & \Psi_{xY}(11,10) & \Psi_{xY}(11,11) 
   \end{array} \right)
   \in \mathbb{R}^{4L \times 4}.
$$
Let $\hat\Psi_{xY}(z,z^\prime) = W^{-1} \sumw M_w^{-1} H_w(z,z^\prime) \hat S_{w,xY(z^\prime)}$. We then have $E(\hat\Psi_{xY}(z,z^\prime)) = \Psi_{xY}(z,z^\prime)$. 

Denote $Q_w = M_w^{-1} (M_w - 1)^{-1} \alphaw^2  \sums (x_{ws} - \bar x_w) \hat Y_{\HT,ws}(z^\prime)$ for $z^\prime = (ab)$ to write
$$\hat\Psi_{xY}(z,z^\prime) = W^{-1} \sumw H_w(z,z^\prime)  Q_w. $$
Similar to the proof above, as $W \rightarrow \infty$, we have
\begin{eqnarray*}
  \cov(\hat\Psi_{xY}(z,z^\prime)) & \leq &  W^{-2}\Big\{p_a \sumw H_w(z,z^\prime)^2 E(Q_w Q_w^\T \mid A_w=a)\Big\} = o(1).
\end{eqnarray*}
Markov’s inequality then ensures that $\hat\Psi_{xY}(z,z^\prime) - \Psi_{xY}(z,z^\prime) = o_\mathbb P(1)$. Therefore, $\hat\Sigma_{*,x\tau} = ( G \otimes I_L) \{ (H\otimes 1_L)\circ(1_4\otimes \hat S_{*,xY}) + \hat\Psi_{xY} \} G^\T = \Sigma_{*,x\tau} + o_\mathbb P(1)$.
\end{proof}

We then introduce a lemma obtained by \citet[][Theorem 2]{zhao2021reconciling}, showing that $\hat\Sigma_{*, \tau\tau}$ is a conservative estimator of $\Sigma_{*, \tau\tau} $ under the classic split-plot randomization. 
\begin{lemma}\label{lem::sample-covariance-yy}
Under Condition~\ref{cond::1}, for $* = \HT,\haj$, 
$$
\hat\Sigma_{*, \tau\tau} - \Sigma_{*, \tau\tau} = G S_* G^\T + o_\mathbb P(1).
$$

\end{lemma}

\begin{proof}[Proof of Theorem~\ref{thm::varest}]
Applying Lemmas~\ref{lem::sample-covariance} and \ref{lem::sample-covariance-yy}, we have
\begin{eqnarray}
             \hat \Sigma_{*} - \Sigma_{*}   = \left(                 
            \begin{array}{cc}   
            GS_{*}G^T & 0_{3 \times 3L}\\  
            0_{3L \times 3} & 0_{3L \times 3L}\\ 
            \end{array}\right)
            + o_{\mathbb P}(1). \nonumber 
\end{eqnarray}
Theorem~\ref{thm::clt} implies that, as $M \rightarrow \infty$, 
$$
\pr( \mathcal{M}_{*}  )  \rightarrow \pr ( \chi^2_{3L} \leq d ) > 0.
$$
Note that if $a_n = o_{\mathbb P}(1)$ then $a_n \mid \mathcal{M}_{*} = o_{\mathbb P}(1)$ because for any $\epsilon >0$,
$$
\pr( | a_n | > \epsilon \mid  \mathcal{M}_{*}) = \pr( | a_n | > \epsilon , \mathcal{M}_{*}) / \pr(\mathcal{M}_{*}) \leq \pr( | a_n | > \epsilon ) / \pr(\mathcal{M}_{*}).
$$
Therefore,
\begin{eqnarray}
            ( \hat \Sigma_{*} - \Sigma_{*} )  \mid \mathcal{M}_{*} = \left(                 
            \begin{array}{cc}   
            GS_{*}G^T & 0_{3 \times 3L}\\  
            0_{3L \times 3} & 0_{3L \times 3L}\\ 
            \end{array}\right)
            + o_{\mathbb P}(1).  \nonumber
\end{eqnarray}
\end{proof}

\subsection{Proof of Corollary~\ref{cor::CI}}
First, we introduce Lemma~\ref{lem::compp_norm} below obtained by \citet[][Lemma A22]{Li2020factorial}.

\begin{lemma}\label{lem::compp_norm}
Let $V_1, V_2 \in \mathbb{R}^{m \times m}$ be two positive semi-definite matrices satisfying $V_1 \leq V_2$, and $\epsilon_1$ and $\epsilon_2$ be two Gaussian random vectors with mean zero and covariance matrices $V_1$ and $V_2$. Then $\epsilon_1 \succ \epsilon_2$.
\end{lemma}

\begin{proof}[Proof of Corollary~\ref{cor::CI}]
Theorem~\ref{thm::varest} ensures that, for $* = \HT, \haj$, $\hat \Sigma_{*,\tau\tau}^{\bot} - \Sigma_{*,\tau\tau}^{\bot} = G S_* G^\T + o_{\mathbb P}(1)$ and $\hat \Sigma_{*,\tau\tau}^{||} - \Sigma_{*,\tau\tau}^{||} = o_{\mathbb P}(1)$. Hence,
$$
    \phi_* \rs (\Sigma_{*,\tau\tau}^{\bot} + G S_* G^\T)^{1/2}\epsilon
            + \Sigma_{*,\tau x}\Sigma_{*,xx}^{-1/2}\zeta_{3L, d}.
$$
By Lemma~\ref{lem::compp_norm}, we have $(\Sigma_{*,\tau\tau}^{\bot} )^{1/2}\epsilon \succ (\Sigma_{*,\tau\tau}^{\bot} + G S_* G^\T)^{1/2}\epsilon$ and $\Sigma_{*,\tau x}\Sigma_{*,xx}^{-1/2}\zeta_{3L, d}$ is, by Lemma~\ref{lemma::ccu}, central symmetric unimodal. This, coupled with Lemma~\ref{lemma::peak2}, ensures that
$$
    (\Sigma_{*,\tau\tau}^{\bot} )^{1/2}\epsilon + \Sigma_{*,\tau x}\Sigma_{*,xx}^{-1/2}\zeta_{3L, d}
    \succ (\Sigma_{*,\tau\tau}^{\bot} + G S_* G^\T)^{1/2}\epsilon + \Sigma_{*,\tau x}\Sigma_{*,xx}^{-1/2}\zeta_{3L, d}.
$$
Recall Theorem~\ref{thm::rerand} and the definition of $\succ$, we have
\begin{eqnarray*}
    \pra(W (\hat \tau_{*} - \tau)^\T (\hat \Sigma_{*,\tau\tau}^{\bot})^{-1} (\hat \tau_{*} - \tau) \leq \hat c_{*, 1 - \xi})
    \geq \pra( \phi_*^\T (\hat \Sigma_{*,\tau\tau}^{\bot})^{-1} \phi_* \leq \hat c_{*, 1 - \xi}) = 1 - \xi.
\end{eqnarray*}


Lemmas~\ref{lemma::peak1} and \ref{lemma::peak3} ensure that $\hat \Sigma_{*,\tau x} \Sigma_{*,xx}^{-1/2} \zeta_{3L, d} \succ \hat \Sigma_{*,\tau x} \Sigma_{*,xx}^{-1/2} D$. Since $(\hat\Sigma_{*,\tau\tau}^{\bot} )^{1/2}\epsilon$ is, by Lemma~\ref{lemma::ccu}, central symmetric unimodal, then by Lemma~\ref{lemma::peak2}, we have 
$$
    \phi_* = (\hat\Sigma_{*,\tau\tau}^{\bot} )^{1/2}\epsilon + \hat\Sigma_{*,\tau x}\Sigma_{*,xx}^{-1/2}\zeta_{3L, d} \succ
    (\hat\Sigma_{*,\tau\tau}^{\bot} )^{1/2}\epsilon + \hat\Sigma_{*,\tau x}\Sigma_{*,xx}^{-1/2}D \sim \hat \Sigma^{1/2}_{*,\tau\tau} \epsilon.
$$
Thus,
\begin{eqnarray*}
    && \pra(\phi_*^\T \hat \Sigma_{*,\tau\tau}^{-1} \phi_* \leq \chi^2_{3, 1 - \xi})
    \geq \pra\{( \hat\Sigma^{1/2}_{*,\tau\tau} \epsilon )^\T \hat \Sigma_{*,\tau\tau}^{-1} (\hat\Sigma^{1/2}_{*,\tau\tau} \epsilon) \leq \chi^2_{3, 1 - \xi}\}
    = 1 - \xi,
\end{eqnarray*}
which suggests $\hat c_{*, 1 - \xi} \leq \chi^2_{3, 1 - \xi}$. Since $  \hat \Sigma_{*,\tau\tau} \geq \hat \Sigma_{*,\tau\tau}^{\bot} $, 
\begin{eqnarray}
  \{ \tau:  W (\hat \tau_{*} - \tau)^\T (\hat \Sigma_{*,\tau\tau}^{\bot})^{-1} (\hat \tau_{*} - \tau) \leq \hat c_{*, 1 - \xi}  \} & \subset & \{ \tau:  W (\hat \tau_{*} - \tau)^\T \hat \Sigma_{*,\tau\tau}^{-1} (\hat \tau_{*} - \tau) \leq \hat c_{*, 1 - \xi}  \} \nonumber \\
  & \subset & \{ \tau:  W (\hat \tau_{*} - \tau)^\T \hat \Sigma_{*,\tau\tau}^{-1} (\hat \tau_{*} - \tau) \leq \chi^2_{3,1-\xi}  \}. \nonumber
\end{eqnarray}
Therefore, the area of the confidence region 
$\{\tau: W (\hat \tau_{*} - \tau)^\T (\hat \Sigma_{*,\tau\tau}^{\bot})^{-1} (\hat \tau_{*} - \tau) \leq \hat c_{*, 1 - \xi}\}$ 
is smaller than or equal to that of the confidence region 
$
\{ \tau:  W (\hat \tau_{*} - \tau)^\T \hat \Sigma_{*,\tau\tau}^{-1} (\hat \tau_{*} - \tau) \leq \chi^2_{3, 1 - \xi}  \}.
$
\end{proof}



\subsection{Proof of Theorem~\ref{thm::rerandreg}}
Define $\Psi_{vY(z)}$ similarly to $\Psi_{xY(z)}$ with $x_{ws}$ replaced by $v_{ws}$. 
Let $T_{vv(z)} = S_{\HT, vv} + p_a \Psi_{vv} $ and $T_{vY(z)} = S_{\HT, vY(z)} + p_a \Psi_{vY(z)}$ for $z=(ab)$. \citet[][Lemma S11 and Proposition 4]{zhao2021reconciling} showed that $\hat \gamma_{\dagger,z}$ has finite probability limits, and linked adjusted estimator $\hat \beta_{\dagger,\L}$ to unadjusted $\hat Y_{*}$; see Lemma~\ref{lemma::Lincoe} below. Let $\hat \beta_{\wls,\L}(z)$ and $\hat \beta_{\ag,\L}(z)$ be the elements in $\hat \beta_{\wls,\L}$ and $\hat \beta_{\ag,\L}$ that correspond to treatment $z$.

Define $S_{*,\L}$ similarly to $S_{*}$ for $*=\HT,\haj$, with $Y_{ws}(z)$ replaced by $Y_{ws}(z; \gamma_{\ag,z})$ and $Y_{ws}(z; \gamma_{\wls,z})$, respectively. Define $\hat v_{*}$ similarly to $\hat x_{*}$ with $x_{ws}$ replaced by $v_{ws}$.
\begin{lemma}\label{lemma::Lincoe}
Under Conditions~\ref{cond::1}--\ref{cond::3}, for $\dagger = \wls, \ag$, $* = \HT, \haj$, and $z = (ab) \in \mathcal{T}$,
\begin{eqnarray*}
&&\hat \gamma_{\dagger, z} = \gamma_{\dagger, z} + o_{\mathbb P}(1), \quad  \hat \beta_{\wls,\L}(z) = \hat Y_{\haj}(z) - \hat v_{\haj}^\T(z) \hat \gamma_{\wls,z}, \quad \hat \beta_{\ag,\L}(z) = \hat Y_{\HT}(z) - \hat v_{\HT}^\T(z) \hat \gamma_{\ag,z}, \\
 && \hat \Sigma_{*,\L, \tau \tau} -  \Sigma_{*,\L, \tau \tau} =  GS_{*,\L}G^T +  o_{\mathbb P}(1),
\end{eqnarray*}
where $\gamma_{\wls, z} = Q_{vv}^{-1}Q_{vY(z)}$, $\gamma_{\ag, z} = T_{vv(z)}^{-1}T_{vY(z)}$, and $S_{*,\L}$ is a positive semi-definite matrix.
\end{lemma}

\begin{proof}[Proof of Theorem~\ref{thm::rerandreg}]
Define $\hat Y_{*}(z; \gamma_{\dagger,z} )$ similarly to $\hat Y_{*}(z)$ with $Y_{ws}(z)$ replaced by $Y_{ws}(z; \gamma_{\dagger,z})$ for $z\in \mathcal{T}$,  where $\dagger = \ag$ for $*=\HT$ and $\dagger = \wls$ for $*=\haj$. Let $\hat Y_{*}( \gamma_{\dagger} )$ vectorize the $\hat Y_{*}(z; \gamma_{\dagger,z} )$'s in lexicographical order of $z$. By Lemma~\ref{lemma::Lincoe} and Theorem~\ref{thm::clt},
$$
\sqrt{W} \{ \hat \beta_{\wls,\L} - \bar{Y} \} =  \sqrt{W} \{ \hat Y_{\haj}(  \gamma_{\wls} ) - \bar{Y}  \} + o_{\mathbb P}(1).
$$
Then
$$
\sqrt{W} ( \hat \tau_{\haj, \L} - \tau ) =  \sqrt{W} \{ G \hat Y_{\haj}( \gamma_{\wls} ) - \tau  \} + o_{\mathbb P}(1).
$$
Since $\pra(\mathcal{M}_{\haj}) = \pr(\chi^2_{3L} \leq d)> 0$, 
$$
\sqrt{W} ( \hat \tau_{\haj, \L} - \tau ) \mid \mathcal{M}_{\haj} =  \sqrt{W} \{ G \hat Y_{\haj}( \gamma_{\wls} )  - \tau  \}  \mid \mathcal{M}_{\haj}  + o_{\mathbb P}(1).
$$
It is straightforward to verify that $Y_{ws}(z; \gamma_{\wls,z})$'s satisfy Condition~\ref{cond::1}. Thus, applying Theorem~\ref{thm::rerand} to $Y_{ws}(z; \gamma_{\wls,z})$, we have
\begin{eqnarray*}
 \sqrt{W}(\hat\tau_{\haj, \L}- \tau) \mid \mathcal{M}_{\haj} &\rs& (\Sigma_{\haj, \L, \tau\tau}^{\bot})^{1/2}\epsilon+\Sigma_{\haj, \L, \tau x}\Sigma_{\haj,  xx}^{-1/2}\zeta_{3L, a}.
\end{eqnarray*}
Lemma~\ref{lemma::Lincoe}, together with $\pra(\mathcal{M}_{\haj}) > 0$, implies 
$$
( \hat \Sigma_{\haj,\L, \tau \tau} -  \Sigma_{\haj,\L, \tau \tau} )  \mid \mathcal{M}_{\haj} =  GS_{\haj,\L}G^T +  o_{\mathbb P}(1).
$$
Applying Theorem~\ref{thm::varest} to $Y_{ws}(z; \gamma_{\wls,z})$, together with Lemma~\ref{lemma::Lincoe}, we have
$$
( \hat \Sigma_{\haj, \L, \tau x} -  \Sigma_{\haj, \L, \tau x} )  \mid \mathcal{M}_{\haj} =  o_{\mathbb P}(1).
$$

The proof for the results regarding the Horvitz–-Thompson estimator $\hat \tau_{\HT,\L}$ is similar, so we omit it.
\end{proof}

\subsection{Proof of Corollary~\ref{cor::varl}}

Lemma~\ref{lemma::Phicond} below comes from \citet[][Lemma S10]{zhao2021reconciling}.

\begin{lemma}\label{lemma::Phicond}
Under Conditions~\ref{cond::1}--\ref{cond::3}, if $\Psi_{vv} = o(1)$, then $\Psi_{vY}  = o(1)$.
\end{lemma}

\begin{proof}[Proof of Corollary~\ref{cor::varl}]
Define $V_{*,vv}$ and $V_{*,vY}=V^\T_{*,Yv}$ similarly to $V_{*,xx}$ and $V_{*,xY}=V^\T_{*,Yx}$ with $x_{ws}$ replaced by $v_{ws}$. Let $V_{*,v(z)Y(z^\prime)} = W \cova ( \hat v_*(z), \hat Y_*(z^\prime))$ denote the asymptotic covariance between $\sqrt{W} \hat v_*(z)$ and $\sqrt{W} \hat Y_*(z^\prime)$, corresponding to the $(z, z^\prime)$ sub-matrix of $V_{*, vY}$. Similarly, let $V_{*,v(z)v(z^\prime)} = W \cova( \hat v_*(z), \hat v_*(z^\prime))$ denote the asymptotic covariance between $\sqrt{W} \hat v_*(z)$ and $\sqrt{W} \hat v_*(z^\prime)$, corresponding to the $(z, z^\prime)$ sub-matrix of $V_{*, vv}$.
For $\dagger = \wls, \ag$, let $\gamma_{\dagger} = \text{diag}(\gamma_{\dagger,00}, \gamma_{\dagger,01}, \gamma_{\dagger,10}, \gamma_{\dagger,11}) \in \mathbb R^{ 4J \times 4}$.
Simple calculation gives
\begin{eqnarray*}
\Sigma_{\HT,\L,\tau x} &=& G ( V_{\HT, Yv} - \gamma_{\ag}^\T V_{\HT,vv} ) (1_{4 \times 4} \otimes C)^\T ( G \otimes I_L)^\T, \\
\Sigma_{\haj,\L,\tau x} &=& G ( V_{\haj, Yv} - \gamma_{\wls}^\T V_{\HT,vv} ) (1_{4 \times 4} \otimes C)^\T ( G \otimes I_L)^\T.
\end{eqnarray*}
For all $z, z^\prime \in \mathcal{T}$, let $ V_{ *,\L,v(z) Y(z^\prime)}  = V_{*, v(z)Y(z^\prime)} - V_{*,v(z) v(z^\prime)}\gamma_{\dagger,z^\prime}$ denote the asymptotic covariance between $\sqrt W \hat v_{*} (z)$ and $\sqrt W \hat Y_{*} (z^\prime, \gamma_{\dagger,z^\prime})$, corresponding to the $(z, z^\prime)$ sub-matrix of $(V_{*, Yv} - \gamma_{\dagger}^\T V_{*,vv})^\T$, where $\dagger = \ag$ for $*=\HT$ and $\dagger = \wls$ for $*=\haj$.


Recall that $\gamma_{\ag, z} = T_{vv(z)}^{-1}T_{vY(z)}$ and $\gamma_{\wls, z} = Q_{vv}^{-1}Q_{vY(z)}$. Let $H(z,z^\prime)$ be the element of $H$ that corresponds to $(z,z^\prime)$. Under $\Psi_{vv} = o(1)$ and Conditions~\ref{cond::1}--\ref{cond::3}, together with Lemma~\ref{lemma::Phicond}, we  have
\begin{eqnarray*}
&& V_{\HT, \L, v(z) Y(z^\prime)} \\
&=&  H(z,z^\prime)S_{\HT, vY(z^\prime)} + o(1) - \{H(z,z^\prime) S_{\HT,vv} + o(1) \} T_{vv(z^\prime)}^{-1}T_{vY(z^\prime)}\\
&=&  H(z,z^\prime)S_{\HT, vY(z^\prime)} + o(1) - \{H(z,z^\prime) S_{\HT,vv} + o(1) \} \{S_{\HT, v v} + o(1)\}^{-1}\{S_{\HT, vY(z^\prime)} + o(1)\}\\
&=& o(1).
\end{eqnarray*}
This ensures that $\Sigma_{\HT,\L,\tau x} = o(1)$. Since $\Sigma_{\HT, \L, \tau\tau}^{||} = \Sigma_{\HT,\L,\tau x} \Sigma_{\HT, xx}^{-1} \Sigma_{\HT,\L,\tau x}^\T = o(1) $, we then have $\Sigma_{\HT, \L, \tau\tau}^{\bot} = \Sigma_{\HT, \L, \tau\tau} - \Sigma_{\HT, \L, \tau\tau}^{||} = \Sigma_{\HT, \L, \tau\tau} + o(1)$. These, together with Theorem~\ref{thm::rerandreg} and Corollary~\ref{cor::varcom}, give
\begin{eqnarray*}
  \sqrt{W}(\hat\tau_{\HT, \L}- \tau) \mid \mathcal{M}_{\HT}   \rs   (\Sigma_{\HT, \L, \tau\tau}^{\bot})^{1/2}\epsilon,\\
 ( { \hat \Sigma}_{\HT,\L, \tau \tau} -  \Sigma^{\bot}_{\HT,\L, \tau \tau} )  \mid \mathcal{M}_{\HT} = GS_{\HT,\L}G^\T +   o_{\mathbb P}(1).
\end{eqnarray*}

From the above results we can derive that
\begin{eqnarray*}
    W\left[\cova(\hat\tau_{\HT} \mid \mathcal{M}_{\HT}) - \cova(\hat\tau_{\HT, \L} \mid \mathcal{M}_{\HT}) \right] &=& \Sigma_{\HT, \tau\tau}^{\bot} - \Sigma_{\HT, \L, \tau\tau}^{\bot} + r_{3L, d}\Sigma_{\HT, \tau\tau}^{||}.
\end{eqnarray*}
By Lemma~\ref{lemma::Lincoe}, we have $\gamma_{\ag,z} = T_{vv(z)}^{-1} T_{vY(z)}$. Condition $\Psi_{vv}=o(1)$, together with the definition of $T_{vv(z)}$ and $T_{vY(z)}$, gives
 $\gamma_{\ag,z} = S_{\HT,vv}^{-1}S_{\HT,vY(z)} + o(1)$. We then have
\begin{eqnarray*}
    & & \Sigma_{\HT, \tau\tau}^{\bot} - \Sigma_{\HT, \L, \tau\tau}^{\bot} \\
    &=&  \Sigma_{\HT, \tau\tau}^{\bot} - \Sigma_{\HT, \L, \tau\tau} + o(1) \\
    &=& \Sigma_{\HT,\tau\tau} - \Sigma_{\HT,\tau x} \Sigma_{\HT,xx}^{-1} \Sigma_{\HT,\tau x}^\T - G (V_{\HT,YY} -\gamma_\ag^\T V_{\HT,vY} - V_{\HT,vY}^\T \gamma_\ag + \gamma_\ag^\T V_{\HT,vv} \gamma_\ag ) G^\T + o(1) \\
    &=& \Sigma_{\HT,\tau\tau} - \Sigma_{\HT,\tau x} \Sigma_{\HT,xx}^{-1} \Sigma_{\HT,\tau x}^\T - \Sigma_{\HT,\tau\tau} + G (\gamma_\ag^\T V_{\HT,vv} \gamma_\ag ) G^\T + o(1)\\
    &=&  - \Sigma_{\HT,\tau x} \Sigma_{\HT,xx}^{-1} \Sigma_{\HT,\tau x}^\T + G (V_{\HT, vY}^\T V_{\HT,vv}^{-1}V_{\HT, vY}   ) G^\T + o(1)\\
    &\geq& - \Sigma_{\HT,\tau v} \Sigma_{\HT,vv}^{-1} \Sigma_{\HT,\tau v}^\T + G (V_{\HT, vY}^\T V_{\HT,vv}^{-1}V_{\HT, vY}   ) G^\T + o(1)\\
    &=& G \Big\{ V_{\HT, vY}^\T \big[ - (G \otimes I_J)^\T \left[(G \otimes I_J)V_{\HT,vv}(G \otimes I_J)^\T\right]^{-1} (G \otimes I_J) + V_{\HT,vv}^{-1} \big] V_{\HT, vY}   \Big\} G^\T + o(1).
\end{eqnarray*}
Note that $V_{\HT, vY}^\T V_{\HT,vv}^{-1} V_{\HT, vY}$ is the covariance of the projection of $\sqrt{W}\hat Y_{\HT}$ on $\sqrt{W}\hat v_{\HT}$, while $V_{\HT, vY}^\T (G \otimes I_J)^\T \left[(G \otimes I_J)V_{\HT,vv}(G \otimes I_J)^\T\right]^{-1} (G \otimes I_J) V_{\HT, vY}$ is the covariance of the projection of $\sqrt{W} \hat Y_{\HT}$ on $\sqrt{W} (G \otimes I_J) \hat v_{\HT}$. Thus, we have $\Sigma_{\HT, \tau\tau}^{\bot} - \Sigma_{\HT, \L, \tau\tau}^{\bot} \geq 0$.
\end{proof}

\subsection{Proof of Theorem~\ref{thm::project}}
\begin{proof}[Proof of Theorem~\ref{thm::project}]
Similar to the proof of Theorem~\ref{thm::rerand}, we denote $\sqrt{W} \tilde\tau_* = (\Sigma_{*,\P, \tau \tau}^\bot)^{1/2} \epsilon + \Sigma_{*,\tau v}\Sigma_{*,vv}^{-1/2} D$ and $\sqrt{W} \tilde \tau_{*,v} = \Sigma_{*,vv}^{1/2}D$, where $\epsilon \sim \mathcal{N}(0, I_{3})$ and $D=(D_1,...D_{3J})^\T \sim \mathcal{N}(0,I_{3J})$ are independent.  Recall that $x_{ws} = C v_{ws}$ for all $(ws) \in \mathcal{S}$. Standard algebra gives $\sqrt{W} \tilde \tau_{*,x} = \sqrt{W} (I_3 \otimes C)  \tilde \tau_{*,v} = (I_3 \otimes C) \Sigma_{*,vv}^{1/2}D $. By Theorem~\ref{thm::clt} and \citet[][Proposition A1]{Li2018},
\begin{eqnarray}
  & & \sqrt{W} 
    ( \hat \tau_* -  \Sigma_{*,\tau v} \Sigma_{*,vv}^{-1} \hat \tau_{*,v} - \tau ) \mid \hat \tau_{*,x}^\T \cov( \hat \tau_{*,x} )^{-1} \hat \tau_{*,x} \leq d \nonumber \\
  & \rs &
  \sqrt{W}
        ( \tilde\tau_* -  \Sigma_{*,\tau v} \Sigma_{*,vv}^{-1} \tilde \tau_{*,v} ) \mid \tilde \tau_{*,x}^\T \cov( \tilde \tau_{*,x} )^{-1} \tilde \tau_{*,x} \leq d \nonumber \\
  & \sim & (\Sigma_{*,\P, \tau \tau}^\bot)^{1/2}\epsilon \mid   \tilde \tau_{*,x}^\T \cov( \tilde \tau_{*,x} )^{-1} \tilde \tau_{*,x} \leq d \nonumber \\
  &  \sim & (\Sigma_{*,\P, \tau \tau}^\bot)^{1/2}\epsilon, \nonumber
\end{eqnarray}
where the last line is due to the independence of $\epsilon$ and $D$.
Note that the above conclusion holds if we replace $\cov(\cdot)$ by $\cova(\cdot)$. Since $(\hat \Sigma_{*,\tau v} - \Sigma_{*,\tau v}) \mid \mathcal{M}_{*} = o_{\mathbb P}(1)$ and $ \sqrt{W} \hat \tau_{*,v} = O_{\mathbb P}(1) $, 
\begin{eqnarray*}
  \sqrt{W}(\hat\tau_{*,\P}- \tau) \mid \mathcal{M}_{*} & \rs &
  \sqrt{W}
        ( \tilde\tau_* -  \Sigma_{*,\tau v} \Sigma_{*,vv}^{-1} \tilde \tau_{*,v} ) \mid \tilde \tau_{*,x}^\T \cov( \tilde \tau_{*,x} )^{-1} \tilde \tau_{*,x} \leq d \\
  & \sim & (\Sigma_{*,\P, \tau \tau}^\bot)^{1/2} \epsilon.
\end{eqnarray*}


Theorem~\ref{thm::varest} ensures that $(\hat \Sigma_{*, \tau \tau } -  \Sigma_{*, \tau \tau }) \mid \mathcal{M}_{*} = GS_*G^\T + o_{\mathbb P}(1)$ for $*=\HT,\haj$. Thus,
\begin{eqnarray*}
&& (\hat \Sigma_{*,\P, \tau \tau}^\bot -  \Sigma_{*,\P, \tau \tau}^\bot )  \mid \mathcal{M}_{*} \\
&=& (\hat \Sigma_{*, \tau \tau} - \hat \Sigma_{*,\tau v}  \Sigma_{*, vv}^{-1} \hat \Sigma_{*, v \tau} - \Sigma_{*, \tau \tau} + \Sigma_{*,\tau v}  \Sigma_{*, vv}^{-1} \Sigma_{*, v \tau}) \mid \mathcal{M}_{*} = GS_{*}G^\T +   o_{\mathbb P}(1). 
\end{eqnarray*}

Moreover, by Theorem~\ref{thm::rerand},
$$
W \cova(\hat\tau_{*} \mid \mathcal{M}_{*}) = 
    \Sigma_{*,\tau \tau}^{\bot}  + r_{3L, d} \Sigma_{*,\tau \tau}^{||}.
$$
Therefore,
$$
W\left[\cova(\hat\tau_{*} \mid \mathcal{M}_{*}) - \cova(\hat\tau_{*,\P} \mid \mathcal{M}_{*}) \right] = 
    \Sigma_{*,\tau \tau}^{\bot} - \Sigma_{*,\P, \tau \tau}^\bot + r_{3L, d} \Sigma_{*,\tau \tau}^{||} \geq 0,
$$
where the last inequality is because $ \Sigma_{*,\tau \tau}^{\bot} - \Sigma_{*,\P, \tau \tau}^\bot \geq 0$ (Note that $\Sigma_{*,\tau \tau}^{\bot} $ and $\Sigma_{*,\P, \tau \tau}^\bot$ are the asymptotic covariances of $\res(\sqrt{W}  \hts \mid \hat\tau_{*, x})$ and $\res(\sqrt{W}  \hts \mid \hat\tau_{*, v})$, respectively, and $\hat\tau_{*, x}$ is a linear transformation of $\hat\tau_{*, v}$).

\end{proof}

\subsection{Proof of Corollary~\ref{cor::comrr}}


Recall that $Q_{\w,vv} = (N-1)^{-1}\sumw (M_w-1) \alpha_w^{-2} S_{w,vv}$ and $\Psi_{vv}=W^{-1}\sumw M_w^{-1} (H_w\otimes S_{w,vv})$. We then have
\begin{eqnarray*}
    \Psi_{vv} &=& O\left( W^{-1} \bar M^{-1} \bar M^{-1} \bar M^2 \sumw  M_w^{-1} (H_w\otimes S_{w,vv})  \right)\\
    &=& O\left( N^{-1} \bar M^{-1}  \sumw  M_w\alpha_w^{-2} (H_w\otimes S_{w,vv}) \right)\\
    &=& O \left( \bar M^{-1} (N-1)^{-1}  \sumw  (M_w-1)\alpha_w^{-2} (H_w\otimes S_{w,vv}) \right).
\end{eqnarray*}
Note that $H_w = O(1)$ by Condition~\ref{cond::1}. We can then derive $\Psi_{vv} = o(1)$ from $Q_{\w,vv} = o(1)$. Thus, $Q_{\w,vv} = o(1)$ is a stricter condition.


\begin{lemma}\label{lemma::Qcond}
Under Condition~\ref{cond::Qcond}, $Q_{\w,vY(z)}  = o(1)$ for $z \in \mathcal{T}$.
\end{lemma}
\begin{proof}[Proof of Lemma~\ref{lemma::Qcond}]
\begin{eqnarray*}
||Q_{\w,vY(z)}||_{\infty} &\leq& (N-1)^{-1} \sumws  || v_{ws}-\bar v_w||_{\infty} | Y_{ws}(z)-\bar Y_w (z)| \\
&\leq& || Q_{\w,vv} ||_{\infty}^{1/2} Q_{\w}(z,z)^{1/2} = o(1).
\end{eqnarray*}
\end{proof}

\begin{proof}[Proof of Corollary~\ref{cor::comrr}]
We first consider the relative efficiency between the Horvitz--Tompson estimator and Hajek estimator under corresponding rerandomization sachems. According to Corollary~\ref{cor::varcom}, for $* = \HT, \haj$, $W\cova(\hat \tau_* \mid \mathcal{M}_{*}) = \Sigma_{*, \tau\tau} - (1 - r_{3L, d})\Sigma_{*,\tau\tau}^{||} $. We then get
\begin{eqnarray}
&& W[\cova(\hat \tau_{\haj}\mid\mathcal{M}_{\haj}) - \cova(\hat \tau_{\HT}\mid\mathcal{M}_{\HT})] \nonumber \\
& = & \Sigma_{\haj, \tau\tau} - \Sigma_{\HT, \tau\tau} - (1 - r_{3L, d}) ( \Sigma_{\haj,\tau\tau}^{||} - \Sigma_{\HT,\tau\tau}^{||}  ) \nonumber \\ 
& = & G\{H\circ (S_\haj - S_\HT)\}G^\T - (1 - r_{3L, d}) (\Sigma_{\haj, \tau x}\Sigma_{\haj,xx}^{-1}\Sigma_{\haj,x\tau} - \Sigma_{\HT, \tau x}\Sigma_{\HT,xx}^{-1}\Sigma_{\HT,x\tau})  \nonumber \\
& = & \Delta \Sigma_{\tau \tau} - (1 - r_{3L, d}) \Delta \Sigma_{\tau\tau}^{||}, \nonumber 
\end{eqnarray}
where $\Delta \Sigma_{\tau \tau} = G\{H\circ (S_\haj - S_\HT)\}G^\T$ and $\Delta \Sigma_{\tau\tau}^{||} = \Sigma_{\haj, \tau x}\Sigma_{\haj,xx}^{-1}\Sigma_{\haj,x\tau} - \Sigma_{\HT, \tau x}\Sigma_{\HT,xx}^{-1}\Sigma_{\HT,x\tau}$.

If $\bar x=0$ and $\bar Y(z)=0$ for all $z$ or $\alphaw=1$ for all $w$, then $S_{\HT} = S_{\haj}$, $S_{\HT,xY} = S_{\haj,xY}$, and $S_{\HT,xx} = S_{\haj,xx}$. Therefore, $\Delta \Sigma_{\tau \tau} = 0$ and $\Delta \Sigma_{\tau\tau}^{||} = 0$.

If $\bar Y_w(z)$ is constant over all $w$, we have $S_\haj = 0_{4\times4}$ and $S_{\haj, xY} = 0_{4L \times 4}$. Hence, $\Delta \Sigma_{\tau \tau}$ is negative semi-definite. If further assume that  $\Psi_{vv} = o(1)$, then $\Psi_{xx} = o(1)$. Thus, $\Psi_{xY}=o(1)$ by Lemma~\ref{lemma::Phicond}, coupled with $S_{\haj, xY} = 0_{4L \times 4}$, ensures that $\Sigma_{\haj, x \tau} = o(1)$. Then, 
\begin{eqnarray*}
W[\cova(\hat \tau_{\haj}\mid\mathcal{M}_{\haj}) - \cova(\hat \tau_{\HT}\mid\mathcal{M}_{\HT})] &=& - \Sigma_{\HT, b, \tau \tau} + (1 - r_{3L, d}) \Sigma_{\HT, b, x \tau}^\T \Sigma_{\HT, b, x x}^{-1} \Sigma_{\HT, b, x \tau} + o(1),
\end{eqnarray*}
where $\Sigma_{\HT, \b, \tau \tau} = G( H\circ S_\HT )G^\T$, $\Sigma_{\HT, \b, x \tau} = ( G \otimes I_L) \{ (H\otimes 1_L)\circ(1_4\otimes S_{\HT,xY})\} G^\T$, and $\Sigma_{\HT, \b, x x} = ( G \otimes I_L ) ( H\otimes S_{\HT,xx} )  ( G \otimes I_L)^\T$. Here, we use subscript ``b" to signify between whole-plot covariances. Define a new outcome $R_{ws}(z) = \bar Y_w(z)$. Let $\tau_R$ be the main effects and interaction for $R_{ws}(z)$ and $\hat \tau_{*,R}$ be its estimator for $*=\HT,\haj$. Theorem~\ref{thm::clt} then ensures that
\begin{eqnarray}
  \sqrt{W} \left( \begin{array}{cc}
        \hat\tau_{\HT,R} - \tau_R  \\
        \hat\tau_{\HT,x}
  \end{array} \right) \rs 
  \mathcal{N} \left( 0, \left(\begin{array}{cc}
        \Sigma_{\HT, \b, \tau\tau} & \Sigma_{\HT, \b, x \tau}^\T\\  
        \Sigma_{\HT, \b, x \tau} & \Sigma_{\HT,\b, xx}\\
  \end{array} \right) \right). \nonumber
\end{eqnarray}
Therefore, $\Sigma_{\HT, \b, \tau \tau} - \Sigma_{\HT,\b, x \tau}^\T \Sigma_{\HT, \b, x x}^{-1} \Sigma_{\HT, \b, x \tau}$ is positive semi-definite. Hence, $\cova(\hat \tau_{\haj}\mid\mathcal{M}_{\haj}) - \cova(\hat \tau_{\HT}\mid\mathcal{M}_{\HT}) \leq 0$.

Similarly, we can prove that $\cova(\hat \tau_{\haj}\mid\mathcal{M}_{\haj}) - \cova(\hat \tau_{\HT}\mid\mathcal{M}_{\HT}) \geq 0$ if $\alphaw \bar Y_w(z)$ is constant over all $w$.

To compare the efficiency between the projection-based Horvitz--Tompson estimator and Hajek estimator under corresponding rerandomization sachems, we can derive
$$
 W[\cova(\hat \tau_{\haj,\P}\mid\mathcal{M}_{\haj}) - \cova(\hat \tau_{\HT,\P}\mid\mathcal{M}_{\HT})] = \Sigma_{\haj,\P, \tau \tau}^\bot - \Sigma_{\HT,\P, \tau \tau}^\bot  = \Delta \Sigma_{\tau \tau} - \Delta \Sigma_{\P, \tau\tau}^{||}
$$
directly from Theorem~\ref{thm::project}. Here, $\Delta \Sigma_{\P, \tau\tau}^{||} = \Sigma_{\haj, \tau v}\Sigma_{\haj,vv}^{-1}\Sigma_{\haj,v\tau} - \Sigma_{\HT, \tau v}\Sigma_{\HT,vv}^{-1}\Sigma_{\HT,v\tau}$. This can be regarded as a special case of $W[ \cova(\hat \tau_{\haj}\mid\mathcal{M}_{\haj}) - \cova(\hat \tau_{\HT}\mid\mathcal{M}_{\HT})]$ with $r_{3L,a} = 0$ and $x_{ws} = v_{ws}$. The proof is thus omitted.


To compare the efficiency between the regression-adjusted Horvitz--Tompson estimator and Hajek estimator under corresponding rerandomization sachems, if assume uniform design, i.e., $\alphaw = 1$ and $M_{wb} = M_{1b}$ for $w=1,\ldots,W$ and $b=0,1$, then by definition, we have $\hat Y_\HT = \hat Y_\haj$ and $\hat x_\HT = \hat x_\haj$. In uniform design, Condition~\ref{cond::Qcond}, together with Lemma~\ref{lemma::Qcond}, ensures that with $z = (ab) \in \mathcal{T}$,
\begin{eqnarray*}
Q_{vv} &=& Q_{\w,vv} + (N-1)^{-1}N/W \sumw \bar v_w \bar v_w^\T 
= S_{\HT,vv} + o(1) = T_{vv(z)} + o(1),\\
Q_{vY(z)} &=& Q_{\w,vY(z)} + (N-1)^{-1}N/W \sumw \bar v_w \{\bar Y_{w}(z) -\bar Y(z)\}\\
&=& S_{\HT,vY(z)} + o(1)= T_{vY(z)} + o(1).
\end{eqnarray*}
Therefore, $\gamma_{\ag, z} = \gamma_{\wls, z} + o(1)$. This, by Lemma~\ref{lemma::Lincoe}, coupled with $\hat Y_\HT = \hat Y_\haj$ and $\hat x_\HT = \hat x_\haj$, ensures that $ W \cova(\hat \tau_{\haj,\L}\mid\mathcal{M}_{\haj}) = W \cova(\hat \tau_{\HT,\L}\mid\mathcal{M}_{\HT}) $.

By Theorem~\ref{thm::rerandreg} and Corollary~\ref{cor::varcom}, if $\Psi_{vv}=o(1)$, we have
    $$W[\cova(\hat \tau_{\haj,\L}\mid\mathcal{M}_{\haj}) - \cova(\hat \tau_{\HT,\L}\mid\mathcal{M}_{\HT})] = \Delta \Sigma_{\L,\tau \tau} - (1 - r_{3L, d}) \Sigma_{\haj, \L, \tau\tau}^{||},$$
where $\Delta \Sigma_{\L,\tau \tau} = \Sigma_{\haj, \L, \tau\tau} - \Sigma_{\HT, \L, \tau\tau}$.

If $\alphaw \bar Y_w(z)$ is constant over all $w$ and $\Psi_{vv} = o(1)$, then $S_\HT = 0_{4\times4}$, $S_{\HT, xY} = 0_{4L \times 4}$, $S_{\HT, vY} = 0_{4J \times 4}$, $\Psi_{vY}=o(1)$ and $T_{vY(z)} = o(1)$, which suggest that $\gamma_{\ag, z} = o(1)$. Thus, $W\cova(\hat \tau_{\HT,\L}\mid\mathcal{M}_{\HT}) = W\cova(\hat \tau_{\HT}) = G \Psi G^\T$. Standard calculation then gives
\begin{eqnarray*}
&& W[\cova(\hat \tau_{\haj,\L}\mid\mathcal{M}_{\haj}) - \cova(\hat \tau_{\HT,\L}\mid\mathcal{M}_{\HT})]\\
&=& G\left( H\circ ( S_\haj + \gamma_\wls^\T (1_{4\times 4}\otimes S_{\haj,vv}) \gamma_\wls - \gamma_\wls^\T (1_4\otimes S_{\haj,vY}) - (1_4\otimes S_{\haj,vY})^\T \gamma_\wls) \right)G^\T\\
&& - (1 - r_{3L, d}) \Sigma_{\haj,\L,\b,x\tau}^\T \Sigma_{\haj,xx}^{-1} \Sigma_{\haj,\L,\b,x\tau},
\end{eqnarray*}
where $\Sigma_{\haj,\L,\b,x\tau} = ( G \otimes I_L) \{ (H\otimes 1_L)\circ(1_4\otimes S_{\haj,xY} + 1_{4\times 4}\otimes (CS_{\haj,vv})\gamma_\wls) \} G^\T$. Again we use subscript ``b" to signify between whole-plot covariances. Define a new outcome $R_{ws}(z) = \bar Y_w(z) - \gamma_{\wls,z}^\T \bar v_w $, and let $\hat \tau_{*,R}$ be the estimators of the main effects and interaction for $R_{ws}(z)$. 
Theorem~\ref{thm::rerand} and Corollary~\ref{cor::varcom} imply that $$W[\cova(\hat \tau_{\haj,\L}\mid\mathcal{M}_{\haj}) - \cova(\hat \tau_{\HT,\L}\mid\mathcal{M}_{\HT})] = W \cova(\hat \tau_{\haj,R}\mid\mathcal{M}_{\haj}) \geq 0.$$

Similarly, we can prove that $W[\cova(\hat \tau_{\haj,\L}\mid\mathcal{M}_{\haj}) - \cova(\hat \tau_{\HT,\L}\mid\mathcal{M}_{\HT})] \leq 0$ if $\bar Y_w(z)$ is constant over all $w$.

\end{proof}

\subsection{Proof of Corollary~\ref{cor::comrrALL}}
Lemma~\ref{lem::Psi-o1} below is obtained from \citet[][Lemma S11]{zhao2021reconciling}. 
\begin{lemma}\label{lem::Psi-o1}
Let $\Psi(z,z^\prime;\gamma)$ be the analog of $\Psi(z,z^\prime)$ with $Y_{ws}(z)$ replaced by $Y_{ws}(z) - \gamma_z^\T v_{ws} $ with $z, z^\prime \in \mathcal{T}$ and arbitrary vectors $\gamma_z$. Under Conditions~\ref{cond::1}--\ref{cond::3} and $\Psi_{vv} = o(1)$, $\Psi(z,z^\prime;\gamma) = \Psi(z,z^\prime) + o(1)$.
\end{lemma}

\begin{proof}[Proof of Corollary~\ref{cor::comrrALL}]
We add a subscript ``$\alpha$" to denote quantities with  the centered whole-plot size factor $(\alphaw - 1)$ included as an additional covariate in the regression. For example, $S_{\HT,\L,\alpha}$, $\Sigma_{\HT,\L,\alpha,\tau \tau}$, $\gamma_{\ag,\alpha}$, and $\hat\gamma_{\ag,\alpha}$,  are analogs of $S_{\HT,\L}$, $\Sigma_{\HT,\L,\tau \tau}$, $\gamma_{\ag}$, and $\hat \gamma_{\ag}$, respectively, with the centered whole-plot size factor $(\alphaw - 1)$ included as an additional covariate in the regression. For $w = 1, \ldots, W$, let
\begin{eqnarray*}
    u_{w} &=& ( G \otimes I_L) \left(\begin{array}{c}
        h(00)\alphaw \bar x_w  \\
        h(01)\alphaw \bar x_w  \\
        h(10)\alphaw \bar x_w  \\
        h(11)\alphaw \bar x_w
    \end{array}\right),
\end{eqnarray*}
where $h(00)=h(01)= (p_0^{-1}-1)^{1/2}$ and $h(10)=h(11)=- (p_1^{-1}-1)^{1/2}$. Standard algebra gives $(W-1)\sumw u_{w} u_w^\T = ( G \otimes I_L ) \{ H\otimes S_{*,xx}\}  ( G \otimes I_L)^\T$ for $*=\HT,\haj$. Thus, under $\Psi_{vv} = o(1)$, we have $\Sigma_{*,xx} = (W-1)\sumw u_{w} u_w^\T + o(1)$.

For $*=\HT,\haj$, let $\eta_{*} = \Sigma_{*, xx}^{-1}(G \otimes I_L) V_{*, xY}$ and $\eta_{*, z}$ denote the column of $\gamma_{*}$ corresponding to treatment $z$.
Let $S_*^\bot = S_* - (1_4 \otimes S_{*,xY})^\T(G \otimes I_L)^\T \Sigma_{*, xx}^{-1} (G \otimes I_L) (1_4 \otimes S_{*,xY})$ for $* = \HT,\haj$. Similarly, define $S_{*,\L}^\bot$ with $Y_{ws}(z)$ replaced by $Y_{ws}(z) - \gamma_{\ag,z}^\T v_{ws}$ and $Y_{ws}(z) - \gamma_{\wls,z}^\T v_{ws}$, respectively, for $*=\HT$ and $*=\haj$. Define $S_{*,\L,\alpha}^\bot$ similarly to $S_{*,\L}^\bot$ with $v_{ws}$ replaced by $(v_{ws}^\T, \alphaw -1)^\T$. 
Denote $\bar c_w = (\bar v_w^\T, \alphaw-1)^\T$, and
\begin{eqnarray*}
    e_{1,w}(z) &=& h(z)\{\alphaw\bar Y_w(z)- \bar Y(z)\} - \eta_{\HT,z}^\T u_w;\\
    e_{2,w}(z) &=& h(z)\{\alphaw\bar Y_w(z)- \alphaw\bar Y(z)\} - \eta_{\haj,z}^\T u_w;\\
    e_{3,w}(z) &=& h(z)\{\alphaw\bar Y_w(z)- \bar Y(z)\} - \eta_{\HT,z}^\T u_w - \theta_z^\T \bar v_w;\\
    e_{4,w}(z) &=& h(z)\{\alphaw\bar Y_w(z)- \alphaw\bar Y(z)\} - \eta_{\haj,z}^\T u_w - \theta_z^\T \bar v_w;\\
    e_{5,w}(z) &=& h(z)\{\alphaw\bar Y_w(z)- \bar Y(z)\} - \eta_{\HT,z}^\T u_w - \theta_z^\T \bar c_w;\\
    e_{6,w}(z) &=& h(z)\{\alphaw\bar Y_w(z)- \bar Y(z)\} - \eta_{\HT,z}^\T u_w - \eta_{c,z}^\T \bar c_w,
\end{eqnarray*}
where $\theta_z$ is an arbitrary vector, $\eta_{c,z}$ is the coefficient from ols fit of $h(z)\{\alphaw\bar Y_w(z)- \bar Y(z)\} - \eta_{\HT,z}^\T u_w$ on $\bar c_w$ over $\{w: w=1,\ldots,W\}$ such that $e_{6,w}(z)$ is the corresponding residual.
Let $e_k(z) = (e_{k,1}(z), \ldots, e_{k,W}(z))^\T$ and  $S_k(z,z^\prime) = (W-1)^{-1}\sumw e_{k,w}(z)e_{k,w}(z^\prime) = (W-1)^{-1} e_k(z)^\T e_k(z^\prime)$, summarized in lexicographical order as $S_k = (S_k(z, z^\prime))_{4\times 4}$ for $k = 1, \ldots, 6$.


Since $x_{ws} = Cv_{ws}$, $\eta_{*,z}^\T u_w$ is a linear combination of $\bar v_w$. Thus, $e_{w,k}(z) - e_{w,6}(z)$ is a linear combination of $\bar c_w$ for all $k = 1, \ldots, 5$. Standard theory of least squares ensures that $\{e_{k}(z) - e_{6}(z)\}^\T e_{6}(z^\prime) = 0$ for all $k = 1, \ldots, 5$, $z,z^\prime \in \mathcal{T}$, and arbitrary vector $\theta_z$. Then
\begin{eqnarray*}
    S_k - S_6 = (W-1)^{-1}\left(\begin{array}{c}
        e_{k-6}(00)^\T  \\
        e_{k-6}(01)^\T  \\
        e_{k-6}(10)^\T  \\
        e_{k-6}(11)^\T  
    \end{array}\right)
    \left(e_{k-6}(00), e_{k-6}(01) , e_{k-6}(10) , e_{k-6}(11) \right) \geq 0,
\end{eqnarray*}
where $e_{k-6}(z) = e_k(z) - e_6(z)$. 
Note that by $\Psi_{vv} = o(1)$ and Lemma~\ref{lem::Psi-o1}, $GS_1G^\T = \Sigma_{\HT,\tau \tau}^\bot + o(1)$; $GS_2G^\T = \Sigma_{\haj,\tau \tau}^\bot + o(1)$; $GS_3G^\T =  \Sigma_{\HT,\L,\tau \tau}^\bot + o(1)$ when $\theta_z = h(z)\gamma_{\ag,z}$; $GS_4G^\T = \Sigma_{\haj,\L,\tau \tau}^\bot + o(1)$ when $\theta_z = h(z)\gamma_{\wls,z}$, and $GS_5G^\T = \Sigma_{\HT,\L,\alpha,\tau \tau}^\bot + o(1)$ when $\theta_z = h(z)\gamma_{\ag,\alpha,z}$.
Thus, for $* = \HT, \haj$, the following inequalities hold as $M \rightarrow \infty$,
$$
    GS_6G^\T \leq \Sigma_{*,\tau \tau}^\bot, \quad GS_6G^\T \leq  \Sigma_{*,\L,\tau \tau}^\bot, \quad GS_6G^\T \leq  \Sigma_{\HT,\L,\alpha,\tau \tau}^\bot.
$$
Since $\Sigma_{*, \P, \tau\tau}^\bot = \Sigma_{*,\tau \tau}^\bot$ when $v_{ws}=x_{ws}$ and $GS_6G^\T \leq \Sigma_{*,\tau \tau}^\bot$ holds for any $x_{ws} = Cv_{ws}$, we also have $GS_6G^\T \leq \Sigma_{*, \P, \tau\tau}^\bot$.

The Frisch--Waugh--Lovell theorem implies that
\begin{eqnarray*}
    \eta_{c,z} &=& \left\{ (W-1)^{-1}\sumw \bar c_w \bar c_w^\T \right\}^{-1}
    \left( (W-1)^{-1}\sumw \bar c_w \left[ h(z)\left\{ \alphaw\bar Y_w(z)- \bar Y(z)\right\} - \eta_{\HT,z}^\T  u_w \right] \right).
\end{eqnarray*}
As the analog of $\hat\gamma_{\ag,z}$, $\hat\gamma_{\ag,\alpha,z} = \hat T_{cc,z}^{-1} \hat T_{cY,z}$, where
\begin{eqnarray*}
    \hat T_{cc,z} &=& W_a^{-1}\sum_{w: A_w=a}\{\hat c_w(z) - \hat c_\HT(z)\}\{\hat c_w(z) - \hat c_\HT(z)\}^\T\\
    &=& W_a^{-1} \sum_{w: A_w=a} \hat c_w(z)\{\hat c_w(z)\}^\T - \hat c_\HT(z)\{\hat c_\HT(z)\}^\T =  (W-1)^{-1}\sumw \bar c_w \bar c_w^\T + o_\mathbb P(1),\\
    \hat T_{cY,z} &=& W_a^{-1}\sum_{w: A_w=a}\{\hat c_w(z) - \hat c_\HT(z)\}\{\alphaw \hat Y_w(z) - \hat Y_\HT(z)\}^\T\\
    &=& W_a^{-1} \sum_{w: A_w=a} \hat c_w(z)\{\alphaw\hat Y_w(z)\}^\T - \hat c_\HT(z)\{\hat Y_\HT(z)\}^\T\\
    &=&  (W-1)^{-1}\sumw \bar c_w \{ \alphaw\bar Y_w(z)- \bar Y(z)\} + o_\mathbb P(1).
\end{eqnarray*}
Therefore, 
\begin{eqnarray*}
    h(z)\gamma_{\ag,\alpha,z} 
    &=& \eta_{c,z} + \left( \sumw \bar c_w \bar c_w^\T \right)^{-1} \left(\sumw \bar c_w \eta_{\HT,z}^\T u_w\right) + o(1).
\end{eqnarray*}

Corollary~\ref{cor::varl} suggests that under $\Psi_{vv} = o(1)$, $\Sigma_{\HT,\L,\alpha,\tau \tau}^\bot = \Sigma_{\HT,\L,\alpha, \tau \tau} + o(1)$. Then if $\theta_z = h(z)\gamma_{\ag,\alpha,z}$, we have
\begin{eqnarray*}
    H \circ S_{\HT,\L,\alpha}^\bot &=& (W-1)^{-1}\sumw e_{5,w}(z)e_{5,w}(z^\prime)\\
    &=& (W-1)^{-1}\sumw \left\{ e_{5,w}(z) + \eta_{\HT,z}^\T u_w \right\}\left\{e_{5,w}(z^\prime) + \eta_{\HT,z^\prime}^\T u_w\right\}.
\end{eqnarray*}
Standard algebra then gives $H \circ S_{\HT,\L,\alpha}^\bot = S_6 + o(1)$, and thus $\Sigma_{\HT,\L,\alpha,\tau\tau}^\bot = G S_6 G^\T + o(1)$. Thus,
\begin{eqnarray*}
    \Sigma_{\HT,\L,\alpha,\tau\tau}^\bot \leq \min\{ \Sigma_{*,\tau\tau}^\bot, 
    \Sigma_{*,\L,\tau\tau}^\bot, 
    \Sigma_{*, \P, \tau\tau}^\bot: 
    * = \HT, \haj \}.
\end{eqnarray*}

Recall that under Conditions~\ref{cond::1}--\ref{cond::3} and $\Psi_{vv} = o(1)$, for $* = \HT,\haj$, 
\begin{eqnarray*}
    \hat \tau_{\HT,\L,\alpha} \mid \mathcal{M}_\HT & \rs & (\Sigma_{\HT,\L,\alpha,\tau\tau}^\bot)^{1/2}\epsilon,\\
    \hat \tau_{\HT,\L} \mid \mathcal{M}_\HT & \rs & (\Sigma_{\HT,\L,\tau\tau}^\bot)^{1/2}\epsilon,\\
    \hat \tau_{\haj,\L} \mid \mathcal{M}_\haj & \rs & (\Sigma_{\haj,\L,\tau\tau}^\bot)^{1/2}\epsilon + \Sigma_{\haj, \L, \tau x}\Sigma_{\haj,  xx}^{-1/2}\zeta_{3L, a},\\
    \hat \tau_{*} \mid \mathcal{M}_* & \rs & (\Sigma_{*,\tau\tau}^\bot)^{1/2}\epsilon + \Sigma_{*,\tau x}\Sigma_{*,  xx}^{-1/2}\zeta_{3L, a},\\
    \hat \tau_{*,\P} \mid \mathcal{M}_* & \rs & (\Sigma_{*,\P, \tau \tau}^\bot)^{1/2}\epsilon,\\
    \hat \tau_{*,\L}  & \rs & (\Sigma_{*,\L,\tau\tau})^{1/2}\epsilon,\\
    \hat \tau_{*}  & \rs & (\Sigma_{*,\tau\tau})^{1/2}\epsilon,\\
    \hat \tau_{*,\P}  & \rs & (\Sigma_{*,\P, \tau \tau}^\bot)^{1/2}\epsilon.
\end{eqnarray*}
Therefore, $\hat \tau_{\HT,\L,\alpha} \mid \mathcal{M}_{\HT}$ is most peaked around $\tau$ among the estimators above.
\end{proof}


\end{document}